\documentclass[10pt,conference,letterpaper]{IEEEtran}
\IEEEoverridecommandlockouts

\usepackage{cite}
\usepackage{amsmath,amssymb,amsfonts}
\usepackage{textcomp}
\usepackage{xcolor}

\usepackage{kotex}
\usepackage{url}
\usepackage{subcaption}
\usepackage{graphicx}
\usepackage{comment}
\usepackage{booktabs}
\usepackage{threeparttable}
\usepackage{float}
\usepackage{tabularx}
\usepackage{hyperref}
\usepackage{tikz}
\usepackage[noend]{algpseudocode}
\usepackage{textcomp}
\usepackage[table, dvipsnames]{xcolor}
\usepackage[normalem]{ulem}
\usepackage{titlecaps}
\usepackage{makecell}
\usepackage{adjustbox}
\usepackage{enumitem}

\usepackage{url}
\usepackage{hyperref}
\hypersetup{hidelinks,colorlinks=true,linkcolor=green!80!black,citecolor=red!70!black,urlcolor=blue!70!black}
\usepackage{bm}
\usepackage{wrapfig}

\usepackage{multirow}

\usepackage{tikz}

\newcommand*\bcircled[1]{\tikz[baseline=(char.base)]{
    \node[shape=circle,fill=black,text=white,inner sep=1pt, line width=1pt] (char) {\bfseries\footnotesize #1};}}

\usepackage[linesnumbered,ruled,vlined]{algorithm2e}
\usepackage{etoolbox,afterpage}

\SetCommentSty{mycommfont}
\SetKwInput{KwInput}{Input}
\SetKwInput{KwOutput}{Output}

\newcommand{\squishlist}{
\begin{list}{$\bullet$}
	{ \setlength{\itemsep}{0pt}      \setlength{\parsep}{-0pt}
		\setlength{\topsep}{4pt}       \setlength{\partopsep}{0pt}
		\setlength{\listparindent}{-2pt}
		\setlength{\itemindent}{-5pt}
		\setlength{\leftmargin}{1em} \setlength{\labelwidth}{0em}
		\setlength{\labelsep}{0.5em} } }
\newcommand{\squishend}{
\end{list}  }

\def\BibTeX{{\rm B\kern-.05em{\sc i\kern-.025em b}\kern-.08em
    T\kern-.1667em\lower.7ex\hbox{E}\kern-.125emX}}

\author{
\IEEEauthorblockN{Bodon Jeong\textsuperscript{1,3}, Hongsu Byun\textsuperscript{1},
Youngjae Kim\textsuperscript{1}, Weikuan Yu\textsuperscript{2}, 
Kyungkeun Lee\textsuperscript{3}, Jihoon Yang\textsuperscript{1}, Sungyong Park\textsuperscript{1,\dag}\thanks{\dag~Corresponding author.}
}
\IEEEauthorblockA{\textsuperscript{1}\textit{Sogang University}, Seoul, Republic of Korea, 
\textsuperscript{2}\textit{Florida State University}, FL, USA,
\textsuperscript{3}\textit{Samsung Electronics Co.} \\
\{bd91jeong, byhs, youkim, yangjh, parksy\}@sogang.ac.kr, wyu3@fsu.edu, 
kavin.lee@samsung.com}
\vspace{-25pt}
}

\newcommand{\dualblade}{\textsc{Dual-Blade}}

\title{
\dualblade{}: Dual-Path NVMe-Direct KV-Cache Offloading
for Edge LLM Inference
\vspace{-0.4cm}
}

\begin{document}
\maketitle
\setlength{\textfloatsep}{5pt plus 1.0pt minus 2.0pt}

\begin{abstract}
The increasing deployment of Large Language Model (LLM) inference on edge AI systems demands efficient execution under tight memory budgets. A key challenge arises from Key–Value (KV) caches, which often exceed available device memory. Although NVMe-based offloading offers scalable capacity, existing file-based designs rely heavily on the kernel page cache, leading to cache thrashing, unpredictable latency, and high software overhead under memory pressure. We present \dualblade{}, a \textit{dual-path KV residency framework} that dynamically assigns KV tensors to either a page-cache path or an NVMe-direct path based on runtime memory availability. The NVMe-direct path bypasses the filesystem by mapping KV tensors to contiguous logical block address (LBA) regions, enabling low-overhead direct storage access. \dualblade{} further incorporates adaptive pipeline parallelism to overlap storage I/O with GPU DMA, improving inference throughput. Our evaluation shows that \dualblade{} substantially mitigates I/O bottlenecks, reducing prefill and decode latency by up to 33.1\% and 42.4\%, respectively, while improving SSD utilization by 2.2$\times$ across diverse memory budgets.
\end{abstract}


\section{Introduction}
\label{sec:introduction}
\vspace{-2pt}

Driven by privacy concerns and the need for low-latency local inference,
LLM serving is increasingly migrating from cloud infrastructure to edge AI systems, including on-premise and embedded deployments~\cite{zheng2025review, pan2025cost, liu2024mobilellm, wang2025intelligent}. 
These edge AI systems are typically single-GPU platforms operating under tight memory budgets (8--32\, GB DRAM), often adopting unified memory where the CPU and GPU share the same DRAM pool~\cite{nvidia_cuda_pg_unified_system_mem}.
Under these constraints, 7--8B parameter models have emerged as the dominant choice~\cite{team2024gemma, jiang2023mistral7b, dubey2024llama}, enabling applications such as AI assistants, long-form document analysis, multimodal search, and log parsing~\cite{zhang2025longreward, bai2024longbench, liu2023visual, zhong2024logparser}.
Unlike cloud datacenters with elastic resources, however, edge AI systems must operate strictly within fixed physical memory limits, making efficient memory management a central challenge for LLM inference.

On edge AI systems, GPU device memory is often near saturation after loading model parameters.
Some systems further spill a portion of model weights into host DRAM~\cite{aminabadi2022deepspeed, sheng2023flexgen}, reducing DRAM slack for runtime states.
Under these conditions, when the KV-cache, which stores per-token key/value vectors for attention to avoid recomputation~\cite{vaswani2017attention}, exceeds GPU memory, it is pushed into the same host DRAM pool~\cite{sheng2023flexgen, lee2024infinigen}, competing with co-located workloads.
The KV-cache size varies dynamically with context length and batch size, and can exhaust both GPU memory and the remaining host DRAM budget, creating a structural bottleneck for modern LLM inference on edge AI systems.

As a result, host DRAM is often insufficient to accommodate the growing KV-cache, making a DRAM--SSD memory-tiering hierarchy for the KV-cache increasingly important. 
Accordingly, many inference serving platforms have begun integrating NVMe SSDs to extend memory capacity~\cite{llama_cpp, cheng2025lmcache, sheng2023flexgen}.
NVMe SSDs provide terabytes of capacity at a fraction of the cost of DRAM, offering a viable solution for hosting massive KV-caches on single-GPU systems~\cite{song2024powerinfer, zhang2025kvswap, sheng2023flexgen, alizadeh2024llm}. 
However, naively mapping these SSDs via standard file systems introduces severe software-stack overheads and I/O inefficiencies, preventing full utilization of device bandwidth.

\vspace{1pt}
\noindent{\textbf{Motivation.}}
A common memory tiering method in LLM inference systems is to use the OS page-cache via file-backed mapping (\texttt{mmap})~\cite{sheng2023flexgen, llama_cpp}.
However, during the \textit{prefill phase}, massive writes exert pressure on the page-cache, triggering synchronous write-back to reclaim space, leading to write stalls.
Furthermore, in the \textit{decode phase}, the auto-regressive pattern of repeatedly reading the accumulated KV-cache causes severe \textit{page-cache thrashing}, where cached pages are evicted before reuse.
As a result, the effective page-cache hit ratio collapses, crippling I/O performance (§\ref{sec:motiv_1}).

Moreover, the software overhead traversing the virtual file system (VFS), filesystem, block layer, and device driver introduces critical latency that hinders full utilization of the NVMe SSD (§\ref{sec:motiv_2}).
This long-standing bottleneck is widely recognized in the storage domain, motivating numerous kernel-bypass solutions to eliminate redundant layers~\cite{kannan2018designing, spdk:online, zhang2018flashshare, joshi2024passthru}.

Another critical issue is the disruption of sequential locality at the SSD device level. Since the Linux multi-queue block layer (\texttt{blk-mq}) distributes requests across multiple NVMe queues, the originally contiguous KV-cache stream becomes fragmented, resulting in an interleaved and non-sequential arrival order at the SSD controller (§\ref{sec:motiv_3}).

\vspace{1pt}
\noindent{\textbf{Key Design and Contributions.}}
The motivation above reveals three fundamental bottlenecks in existing KV-cache offloading designs:
(i) page-cache thrashing caused by indiscriminate \texttt{mmap}-based placement,
(ii) excessive software overhead along the kernel storage stack, and
(iii) loss of sequential locality at the NVMe device level.
\dualblade{} directly addresses these challenges through three complementary designs:

\squishlist
\item
\textbf{Dual-Path KV Residency.}
\dualblade{} decouples the I/O path into a page-cache path and an NVMe-direct path. KV tensor groups that fit in the dynamically estimated page-cache budget are placed on the page-cache path to avoid thrashing and sustain high hit ratios, while remaining tensors are routed to the NVMe-direct path for direct access to contiguous NVMe logical block addresses (LBAs) (§\ref{section:design_1}).

\item
\textbf{NVMe-direct with Sequential-LBA Placement.} 
To minimize software overhead and maximize SSD utilization, \dualblade{} bypasses the kernel storage stack on the NVMe-direct path. It further enforces sequential LBA placement for KV tensors, preserving device-level sequentiality and reducing Flash Translation Layer (FTL) and controller scheduling overhead (§\ref{section:design_2}).

\item
\textbf{Adaptive Pipeline Parallelism.}
In addition, \dualblade{} incorporates adaptive pipeline parallelism to overlap KV-cache storage I/O with GPU DMA. By dynamically coordinating I/O and computation across threads under bandwidth contention, this design effectively hides residual I/O latency and maximizes end-to-end inference throughput (§\ref{section:design_3}).
\squishend

We implemented the proposed design on a modern disk-offloading-based inference system, FlexLLMGen~\cite{sheng2023flexgen}. The evaluation results demonstrate that our approach reduces end-to-end inference latency by up to 33.1\% in the prefill phase and 42.4\% in the decode phase compared to baseline methods, while boosting SSD utilization by up to 2.2$\times$ and maintaining robust performance under diverse memory constraints.

\begin{figure}[t]
	\centering
    \includegraphics[width=0.9 \linewidth]{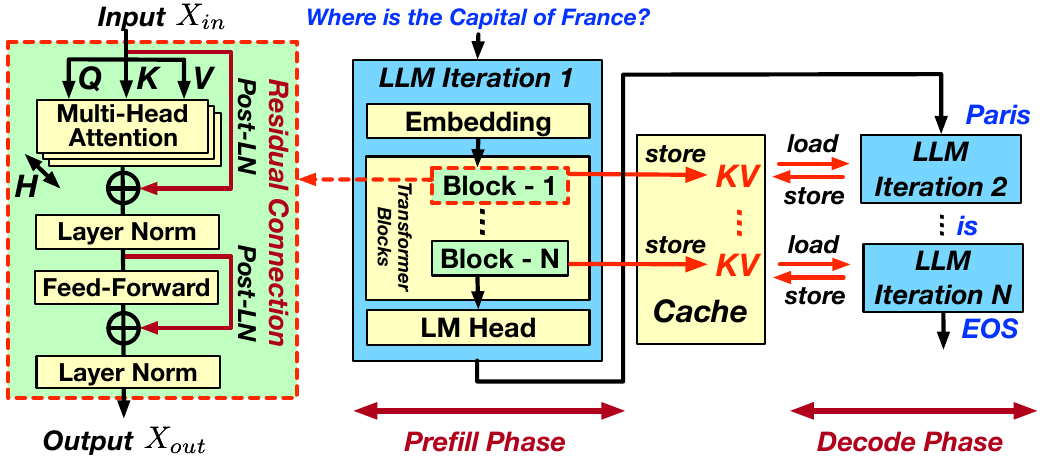}
    \vspace{-3pt}
    \caption{\small LLM transformer architecture~\cite{vaswani2017attention}.}
    \label{fig:llm_transformer_archi}
    \vspace{-3pt}
\end{figure}

\section{Background}
\label{sec:background}
\vspace{-2pt}
\subsection{LLM Transformer Inference and KV-Cache}
\vspace{-2pt}
\label{sec:llm-inference}
Figure~\ref{fig:llm_transformer_archi} shows an LLM built from stacked transformer blocks. Each block has two sublayers, multi-head attention (MHA) and a feed-forward network (FFN), each followed by a residual connection and layer normalization. MHA maps input $X$ to Query ($Q$), Key ($K$), and Value ($V$), computes attention from $Q$--$K$ similarity, and weights $V$ to form a contextual representation for each token representation. The computation runs in parallel over $H$ heads, and their outputs are combined to produce the MHA result. This then goes to the FFN, yielding the block output $X_{out}$ for the next block.

LLM inference has two phases, \textit{Prefill} and \textit{Decode}. In prefill, all prompt tokens pass through the embedding, transformer stack, and language modeling head to produce the first token (\textit{LLM Iteration~1}). That token becomes the input to decode. Decode proceeds in an auto-regressive manner, generating the next token from the previous one (\textit{LLM Iteration~2$\ldots$N}). Generation stops at a maximum length or an EOS token.

During each decode step, the model \textit{loads} all prior KV pairs from the KV-cache to compute attention for the current token, then \textit{stores} the newly generated KV pair back into the KV-cache. This avoids recomputing past KVs at every step, making KV caching a standard technique in modern LLM inference. However, KV-cache size depends on context length and batch size and can readily exceed GPU memory.

\subsection{Memory Tiering for Efficient KV-Cache Management}
\vspace{-2pt}
When KV-cache exceeds GPU memory, LLM inference systems offload it to host DRAM and, increasingly, to NVMe SSDs as an extended memory tier. 
To illustrate KV-cache offloading in practice across diverse inference platforms, we describe FlexLLMGen~\cite{sheng2023flexgen}, a recent LLM serving system that enables high-throughput single-GPU inference by tiering the KV-cache between host DRAM and NVMe SSDs.

\begin{figure}[t]
	\centering
    \includegraphics[width=0.9 \linewidth]{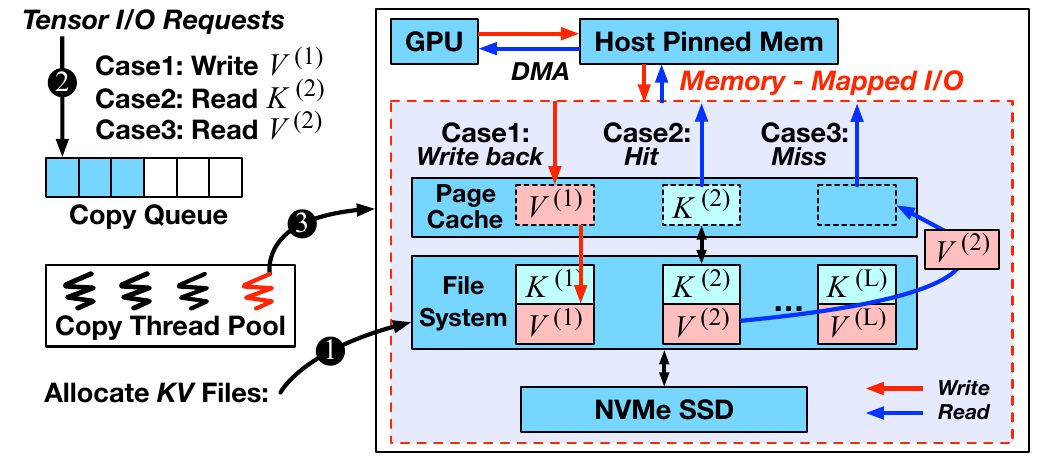}
    \vspace{-3pt}
    \caption{\small KV-cache disk offload workflow~\cite{sheng2023flexgen}.}
    \label{fig:kv_cache_disk_offload_arch}
    \vspace{-3pt}
\end{figure}

\newcommand{\Ki}{K^{(i)}}
\newcommand{\Vi}{V^{(i)}}

As shown in Figure~\ref{fig:kv_cache_disk_offload_arch}, the system first creates \(2L\) files for \(\Ki\) and \(\Vi\) tensors\footnote{Here, ``tensor'' denotes the per-layer KV component (\(\Ki\) or \(\Vi\)). We use the terms interchangeably.} per layer \(i\in\{1,\dots,L\}\) (\bcircled{1}). When a layer is needed during prefill or decode, an I/O request for its \(\Ki/\Vi\) is enqueued to the \textit{Copy Queue} (\bcircled{2}), and a \textit{Copy Thread Pool} worker performs the transfer (\bcircled{3}). Data moves in two segments: GPU\(\leftrightarrow\)CPU via pinned-memory DMA, and CPU\(\leftrightarrow\)filesystem via memory-mapped I/O, with the OS page-cache tiering DRAM and SSD. Writes land in the page cache and flush asynchronously (\textit{Case1: Write \(V^{(1)}\)}). For reads, page-cache hits return immediately (\textit{Case2: Read \(K^{(2)}\)}). On a miss, the kernel fetches from disk, fills the page-cache, and then serves the data (\textit{Case3: Read \(V^{(2)}\)}).

\subsection{Memory Constraints in Edge LLM Serving}
\vspace{-2pt}
\label{sec:memory_constraints_in_edge}
In edge AI systems, GPU memory is largely consumed by model parameters. The KV-cache size varies with context length and batch size, and can exceed GPU memory capacity.

\vspace{1pt} \noindent\textbf{Target Scenario.}
We focus on deployments where the KV-cache outgrows GPU memory, spills into host memory, and is eventually offloaded to SSDs when the host-memory budget is tight under contention from co-located workloads.
To evaluate disk offloading under controlled yet realistic memory pressure, we fix the KV-cache workload and sweep the host-memory limit, spanning regimes from severe under-provisioning (relative to the KV-cache) to ample headroom.

\begin{figure}[t]
    \subfloat[\small Inference latency]{%
    \vspace{-3pt}
    \includegraphics[width=0.46\linewidth]{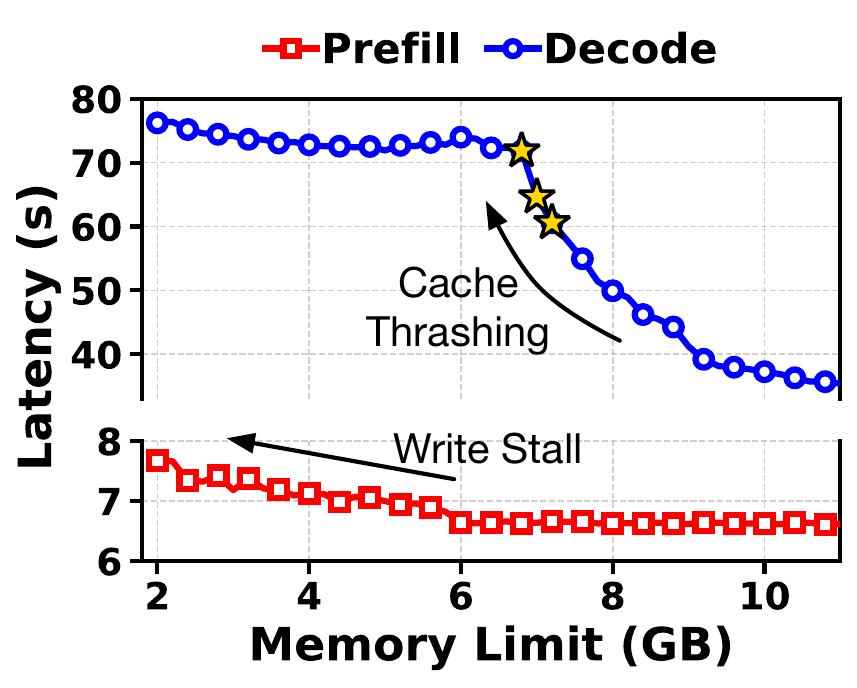}}
    \hfil
    \subfloat[\small Page-cache utilization]{%
    \vspace{-3pt}
    \includegraphics[width=0.5\linewidth]{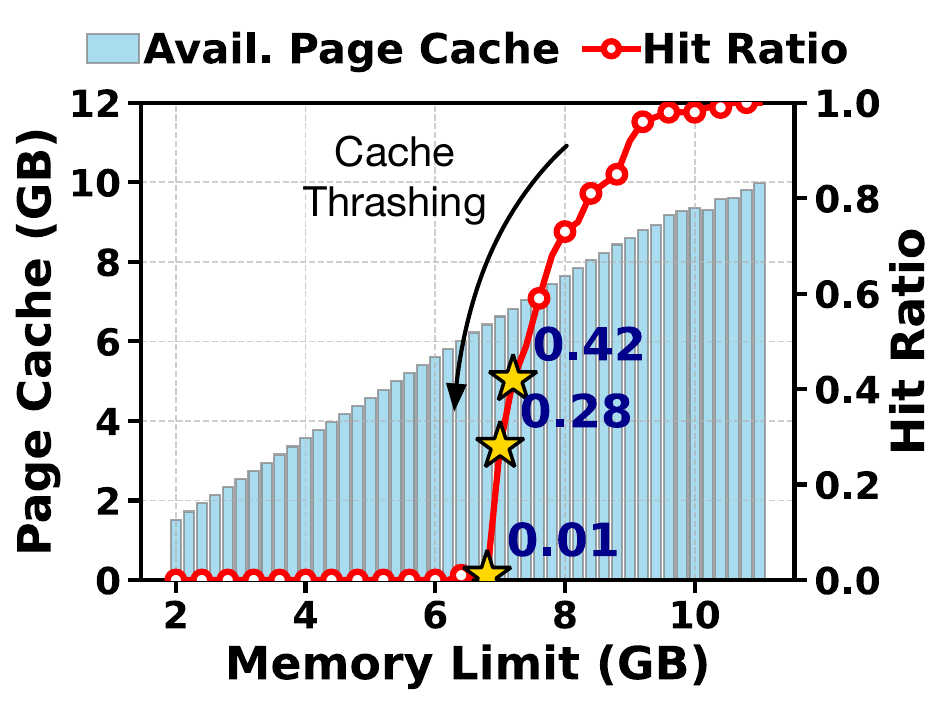}}
    \hfil
    \vspace{-5pt}
    \caption{\small Page-cache thrashing under host memory limits.}
	\vspace{-3pt}
\label{fig:page_cache_thrashing}
\end{figure}

\section{Preliminary Study and Motivation}
\label{sec:motivation}

\subsection{Page-Cache Thrashing in KV Workloads}
\vspace{-2pt}
\label{sec:motiv_1}
We analyze \textit{page-cache thrashing under limited host memory}.
Figure~\ref{fig:page_cache_thrashing} reports prefill/decode latency, available page-cache, and page-cache hit ratio as we vary the host memory limit from 2 to 11\,GB. 
We use OPT-6.7B~\cite{facebook_opt6.7b} with a 512-token prompt, 32-token generation, and batch size 32.
The resulting KV-cache totals \(8.57\)--\(9.11\,\mathrm{GB}\). 
For example, at a host memory limit of 11\,GB, the page-cache can hold the entire KV-cache.
We estimate available page-cache from cgroup stats sampled every 1\,s. The page-cache hit ratio in the decode phase is the fraction of total read bytes served from the page-cache.
The full experimental setup is described in \S\ref{sec:expr_setup}.

Prefill is write-heavy. The monotonic latency rise in Figure~\ref{fig:page_cache_thrashing}(a) once memory drops below \(\approx 6\,\mathrm{\text{GB}}\) follows directly from page-cache behavior. A smaller page-cache forces synchronous evictions before background write-back, stalling write I/O and increasing prefill latency.

Decode is read-heavy. For each generated token, the system rereads the entire accumulated KV-cache. Thus, the reusable working set starts right after prefill and increases each iteration as decode appends new KVs.
With 10--11\,GB, the page-cache can hold the full KV-cache, yielding $\approx 100$\% hit ratio and low latency. As the host memory limit falls below the KV working set (\(8.57\)--\(9.11\,\mathrm{\text{GB}}\)), one might expect the page-cache hit ratio to decline gradually as misses increase. Instead, we observe a sharp cliff. At the starred points in Figure~\ref{fig:page_cache_thrashing}(b), the page-cache hit ratio collapses from 42\% to $<1$\%, coinciding with the spike in decode latency in Figure~\ref{fig:page_cache_thrashing}(a).

This creates a thrashing zone at 2--7\,GB. The hit ratio stays $<1$\% despite several GB of page-cache being available. To our knowledge, this is the first experimental identification of page-cache thrashing as a major bottleneck in KV offloading for LLM inference, consistent with OS literature~\cite{johnson1994x3, megiddo2003arc, li2024streamcache}.

\vspace{1pt}
\noindent\textbf{Root Cause Analysis.}
The root cause is the \textit{conflict between auto-regressive reuse and LRU-based page caching}. 
In decode, attention repeatedly reads the growing KV working set. When the page-cache is smaller than this working set, LRU triggers cascading evictions. Reading one tensor evicts another that will soon be reread, leaving the cache full yet ineffective and driving sustained disk I/O.
Hence, a naive \texttt{mmap}-based design fails to properly tier KV data across DRAM and SSD and can collapse into page-cache thrashing, motivating strategies that structurally avoid thrashing for such KV access patterns.

\begin{figure}[t]
	\centering
    \includegraphics[width=0.96 \linewidth]{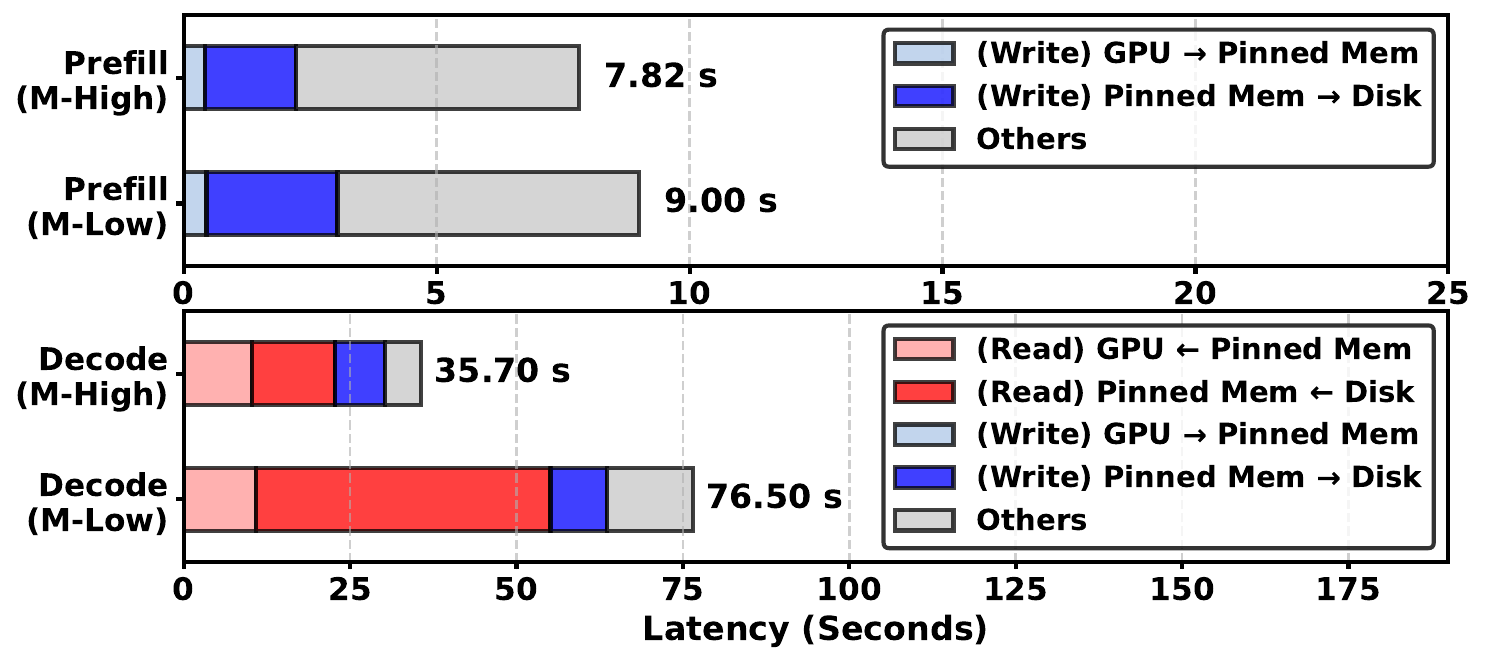}
    \vspace{-5pt}
    \caption{\small Total latency breakdown for prefill and decode phases.}
    \label{fig:sub1_lat_breakdown}
    \vspace{-3pt}
\end{figure}

\subsection{Kernel Storage I/O Path as the Core Inference Bottleneck} 
\vspace{-2pt}
\label{sec:motiv_2}

Figure~\ref{fig:sub1_lat_breakdown} contrasts two extremes of page-cache efficacy, M-High and M-Low, instantiated using the 11\,GB and 2\,GB host-memory limits in Figure~\ref{fig:page_cache_thrashing}. Beyond the obvious benefit that more page-cache hits are better, it shows that \textit{prefill} and \textit{decode} have fundamentally different dominant costs. Here, disk I/O refers to file-backed \texttt{mmap} I/O via the page-cache.

Prefill is a one-shot \textit{large write} of input-token KVs. The performance difference between the two prefill cases mainly comes from the differing impact of page-cache write stalls. In M-High and M-Low, ``Others'' (mostly GPU compute) dominates (72\% and 66\%), indicating that prefill is largely compute dominated. However, disk I/O (``Pinned Mem $\rightarrow$ Disk'') still accounts for a non-trivial fraction (24\% and 28\%).

Decode repeatedly performs \emph{large read + small write}. The performance gap between the two decode cases is driven by their page-cache hit ratios, which are ($\approx 100$\%) in M-High and ($\approx $0\%) in M-Low. In M-High and M-Low, decode accounts for 82\% and 89\% of end-to-end inference time. Within decode, disk I/O (``Pinned Mem $\leftrightarrow$ Disk'') accounts for 56\% (M-High) and 69\% (M-Low), indicating that decode is largely disk I/O dominated rather than compute dominated.

Overall, disk I/O is a major cost in both prefill and decode across the two extreme memory regimes (M-High and M-Low), although it is secondary to GPU compute in prefill. This motivates optimizing the storage I/O path.

\begin{figure}[!h]
	\vspace{-8pt}
    \centering
    \includegraphics[width=0.96 \linewidth]{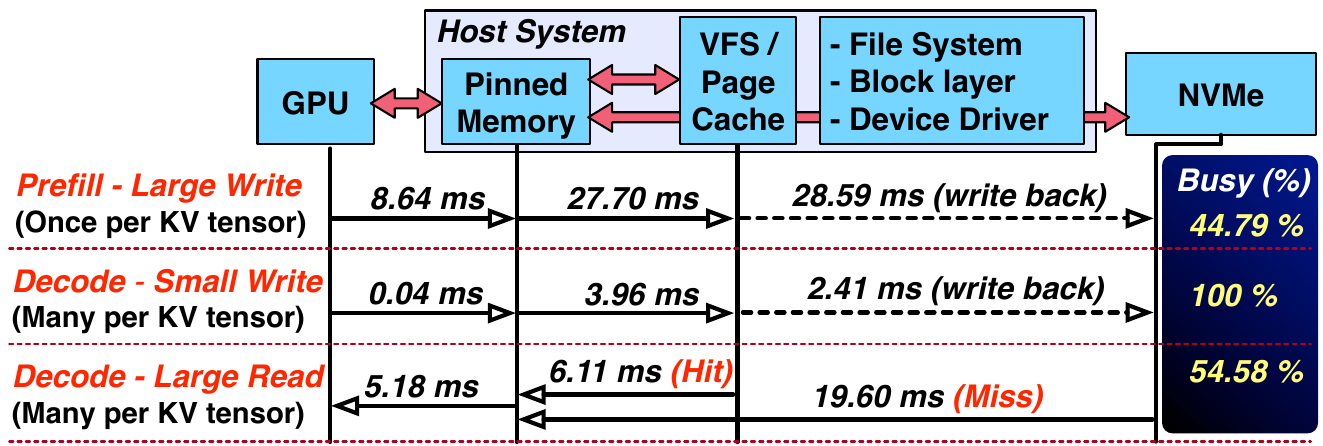}
    \vspace{-3pt}
    \caption{\small Per-tensor I/O request latency breakdown for a single layer.}
    \label{fig:sub2_lat_breakdown}
    \vspace{-5pt}
\end{figure}

Figure~\ref{fig:sub2_lat_breakdown} shows the latency breakdown for each K or V tensor I/O request issued within a single transformer layer.
Request sizes are as follows in our setup: 128\,MB (prefill write), 128--135\,MB (decode read), and 256\,KB (decode write).
We measure the NVMe \emph{busy ratio} (utilization), the fraction of time the device is actively processing I/O, using \texttt{bpftrace}~\cite{gregg2019bpf} at the block layer by correlating submission and completion events and accounting for overlaps and idle gaps across requests over the measurement window.

On write-back or a page-cache miss, requests traverse the full kernel software stack (VFS\(\to\)filesystem\(\to\)block layer\(\to\)driver), accumulating latency that is substantial on a per-request basis.
Decode write (256\,KB) reaches 100\% busy because the kernel issues it as a single I/O (e.g., one \texttt{bio}). In contrast, prefill writes (128\,MB) and decode reads (128--135\,MB) are chunked at the block layer. Each chunk pays full-stack overhead, leaving device idle gaps between commands and yielding low busy ratios (45\% and 55\%)~\cite{zhang2018flashshare, lee2019asynchronous, woo2021d2fq}. 
This underutilizes NVMe hardware. Eliminating per-chunk overhead via a lighter I/O path to maximize device busy time is therefore a key optimization target.

\subsection{Logical-to-Physical Access Disparity in KV-Cache I/O}
\vspace{-2pt}
\label{sec:motiv_3}

The KV-cache workload is flash-friendly. Logically (from the tensor view), it is dominated by large sequential writes (prefill) and large sequential reads (decode).

\begin{figure}[!h]
  \centering
  \vspace{-10pt}

  \captionsetup[subfigure]{labelformat=empty}
  \subfloat[]{%
    \includegraphics[width=0.90\linewidth]{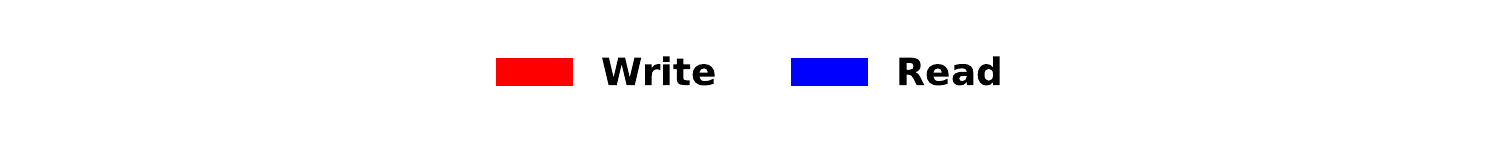}}
  \vspace{-20pt}

  \hspace{-7pt}
  \includegraphics[width=0.275\linewidth]{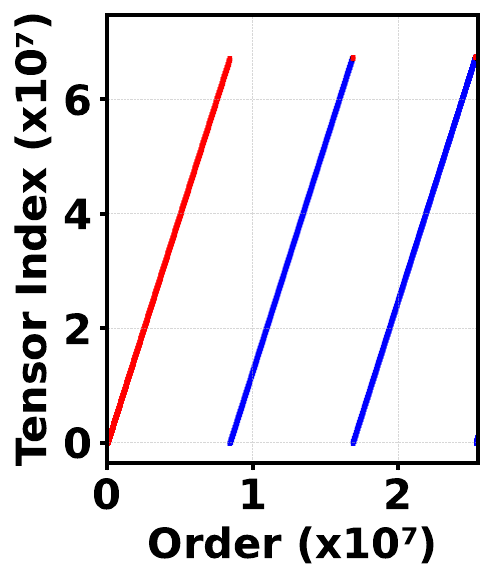}
  \hspace{-7pt}
  \includegraphics[width=0.385\linewidth]{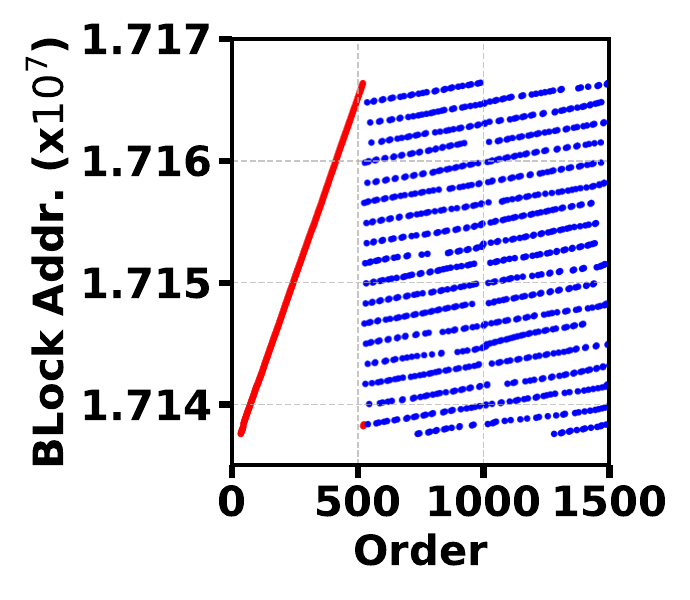}
  \hspace{-7pt}
  \includegraphics[width=0.335\linewidth]{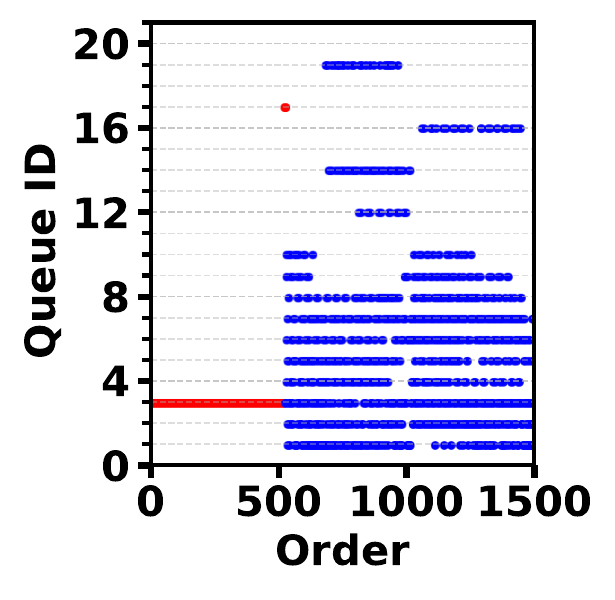}

  \vspace{-5pt}
  
  \hspace{-7pt}
  \makebox[0.275\linewidth]{\small{(a) Tensor Access}}
  \hspace{-7pt}
  \makebox[0.385\linewidth]{\small{(b) LBA Access}}
  \hspace{-7pt}
  \makebox[0.335\linewidth]{\small{(c) Submission Queue}}

  \vspace{-2pt}
  \caption{\small Logical (tensor-level) and physical (device-level) access patterns. The x-axis denotes the I/O index in submission order.}
  \vspace{-5pt}
  \label{fig:motiv_3}
\end{figure}

Figure~\ref{fig:motiv_3}(a) illustrates the linearized logical address over tensor access order, exhibiting near-perfect sequentiality in both the prefill and decode phases.
Decode writes are minimal and appear only as sparse points.
In contrast, the device-level logical block address (LBA) stream breaks this sequentiality, as shown in Figure~\ref{fig:motiv_3}(b), where prefill writes remain strictly increasing as an idealized sequential pattern, whereas decode reads split into multiple interleaved streams.

\vspace{1pt}
\noindent\textbf{Root Cause Analysis.}
The root cause appears in the NVMe submission-queue (SQ) ID distribution of Figure~\ref{fig:motiv_3}(c). Writes concentrate in a few queues driven by a small set of kernel threads~\cite{park2017ijournaling}, whereas reads fan out across many queues. Page-cache misses trigger parallel submissions from application and kernel threads~\cite{bhattacharya2003asynchronous}. Thus a logically single sequential read becomes, after the Linux multi-queue block layer (\texttt{blk-mq}), an interleaved mix of short sequential fragments.

While \texttt{blk-mq} schedulers improve throughput for complex I/O mixes, their kernel-stack and scheduling overhead (on the \(\mu\)s scale) on low-latency NVMe SSDs can be counterproductive for a single sequential stream~\cite{zhang2018flashshare, whitaker2023we, ren2024bfq}. Multi-queueing is generally beneficial, but domain-specific sequential workloads such as KV-cache access can saturate the device with a single thread. Preserving end-to-end sequentiality to the NVMe controller reduces internal work (e.g., FTL mapping, reordering). As shown in \S\ref{sec:motiv_2}, when I/O dominates, shaving even \(\mu\)s-level overhead accumulates and reduces total inference time. Maintaining a strictly sequential submitted-LBA pattern is therefore a sound optimization goal.

\begin{table}[t]
\scriptsize{}
\setlength{\tabcolsep}{0pt}
\renewcommand{\arraystretch}{1}

\centering
\caption{\small I/O Path Comparison (128KB Seq. Read/Write, QD=32)}
\vspace{-4pt}
\label{tab:io-latency}

\begin{tabularx}{\linewidth}{@{} l >{\centering\arraybackslash}X >{\centering\arraybackslash}X >{\centering\arraybackslash}X @{}}
\toprule
\textbf{Metric} & \textbf{ext4 File I/O} & \textbf{\texttt{io\_uring\_cmd}} & \textbf{SPDK} \\
\midrule

\textbf{Path Depth} & 
Deep (VFS/ext4) & 
\textbf{Short} (Kernel-bypass) & 
Short (User Driver) \\ 
\midrule

\textbf{Mechanism} & 
Standard File I/O & 
\textbf{Passthrough} & 
User-space Polling \\ 
\midrule

\begin{tabular}[c]{@{}l@{}} \textbf{Write Latency} \\ \scriptsize{(Avg. / 99.99th)} \end{tabular} & 
668 $\mu$s / 3.50 ms & 
\textbf{662 $\mu$s / 1.62 ms} & 
662 $\mu$s / 1.14 ms \\ 
\midrule

\begin{tabular}[c]{@{}l@{}} \textbf{Read Latency} \\ \scriptsize{(Avg. / 99.99th)} \end{tabular} & 
351 $\mu$s / 17.8 ms & 
\textbf{340 $\mu$s / 1.19 ms} & 
339 $\mu$s / 0.86 ms \\ 
\bottomrule
\end{tabularx}
\end{table}

\subsection{Opportunity: NVMe-direct Path (Kernel-bypass)}
\vspace{-2pt}
The KV-cache workload for LLM inference is fundamentally different from general file I/O. First, its access pattern is predictable and largely sequential. The prefill phase involves sequential writes, while the decode phase requires sequential reads with appended writes. There are no overwrites and tensors are not shared concurrently. Moreover, the inference stack can predetermine the entire KV footprint and on-disk layout before execution.
Second, the KV-cache is temporal and ideally resides in DRAM. Consequently, standard filesystem guarantees such as persistence, consistency, and metadata durability are unnecessary. Leveraging these characteristics, we employ a lightweight mechanism called NVMe-direct that bypasses the filesystem to handle offloaded KV data efficiently.

In our implementation, we use the \emph{NVMe-direct path} as a kernel-bypass channel that issues device-native NVMe commands via \texttt{io\_uring\_cmd}~\cite{joshi2024passthru}. 
To justify this, we compare it with SPDK~\cite{spdk:online} and standard filesystem I/O in Table~\ref{tab:io-latency}. As shown, both bypass methods significantly outperform the filesystem, particularly in tail latency. While SPDK offers slightly lower latency, it incurs high complexity and resource consumption due to busy-polling and core binding. In contrast, \texttt{io\_uring\_cmd} achieves comparable performance through a standard kernel interface. Thus, we adopt \texttt{io\_uring\_cmd} for its balance of performance and ease of implementation, noting that our design is extensible to SPDK.

\section{Design and Implementation}
\vspace{-2pt}

From $\S$\ref{sec:motivation}, KV-cache offloading faces three bottlenecks: (i) page-cache thrashing, (ii) kernel-stack overhead, and (iii) tensor–LBA mismatch. \dualblade{} is a high-performance offloading architecture for edge AI systems under tight memory budgets that addresses them with the following components.

\squishlist
\item
\noindent\textbf{Dual-Path KV Residency.} 
Partition KV tensors into a page-cache path (Group~1) and an NVMe-direct path (Group~2) that bypasses the kernel storage stack. This separation avoids page-cache thrashing during decode, while also preventing write stalls during prefill.

\item
\noindent\textbf{NVMe-direct with Sequential-LBA Placement.} 
Accelerate the NVMe-direct path by issuing device-native NVMe I/O that bypasses the kernel stack, aiming to (i) increase NVMe utilization by eliminating software overhead, and (ii) reduce NVMe controller latency by enforcing a strictly sequential LBA submission pattern.

\item
\noindent\textbf{Adaptive Pipeline Parallelism.}
To further maximize inference performance, in multi-threaded configurations where storage bandwidth becomes saturated, dynamically enforce a staggered start on trailing threads to overlap storage I/O and GPU DMA, thereby mitigating resource contention.
\squishend

\begin{figure}[t]
	\centering
    \includegraphics[width=0.9 \linewidth]{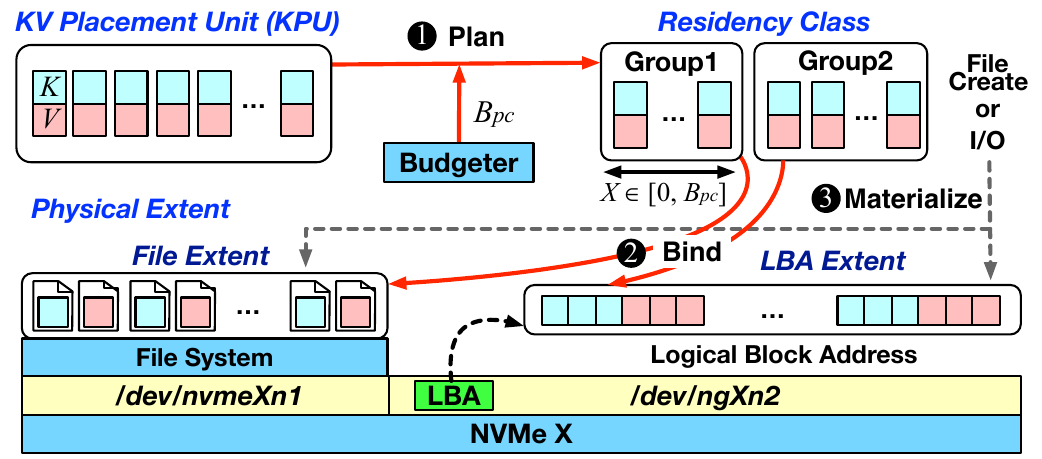}
    \vspace{-3pt}
    \caption{\small High-level architecture of dual-path KV cache residency.}
    \label{fig:dual_path_kv_residency}
\end{figure}

\subsection{Dual-Path KV Residency}
\vspace{-2pt}
\label{section:design_1}
The dual-path KV residency assigns KV tensors of hidden layers to two I/O path groups. Consequently, during the prefill and decode phases, each tensor is processed via the I/O stack corresponding to its assigned path.

Figure~\ref{fig:dual_path_kv_residency} outlines this workflow.
At initialization, each KV pair is abstracted as a \emph{KV Placement Unit} (KPU).
A budgeter computes the usable page-cache and classifies each KPU into two classes: Group~1 (page-cache path) and Group~2 (NVMe-direct path) (\bcircled{1}). 
Each group is then bound to a physical address space, mapping Group~1 to filesystem extents and Group~2 to NVMe-namespace LBA extents (\bcircled{2}).
This binding remains logical until the first access, at which point it is materialized as an actual file or block (\bcircled{3}).
The details of the three steps, \textbf{Plan}, \textbf{Bind}, and \textbf{Materialize}, are as follows.

\vspace{1pt}
\noindent\textbf{Plan.}
First, the budgeter estimates the available page-cache budget $B_{pc}$ under system and cgroup limits:
{
\setlength{\abovedisplayskip}{2pt}%
\setlength{\belowdisplayskip}{2pt}%
\begin{flalign}
& M^* \leftarrow \min(M_{avail},\,M_{max} - M_{anon+shmem}) && \label{eq:mstar}\\[-2pt]
& B_{pc} \leftarrow \max(0,\,M^* - N_{threads} \cdot M_{pin}) && \label{eq:bpc}
\end{flalign}
}
Pinned memory ($M_{pin}$) is reserved exclusively for GPU DMA usage by each copy-thread, matching the size of a single K or V tensor (KPU) of a hidden layer. Consequently, the total reservation $N_{threads} \cdot M_{pin}$ constitutes a constant DRAM overhead for inference serving, distinct from the page-cache.

Algorithm~\ref{alg:classify_kpu_to_residency_class} uses a knob \(X\in[0,B_{pc}]\) to partition each hidden layer’s KPU pair \((K^{(i)},~V^{(i)})\) into Group~1 or Group~2. 
The knob \(X\) sets the total bytes admitted to the page-cache path. 
We first compute how many layers \(n_1\) fit (line~1), then assign KPU pairs from layer~1 to layer~\(n_1\) to Group~1 (\(x_i{=}1\)) and the rest to Group~2 (\(x_i{=}0\)) (lines~2--4).

Prior work ranks KV entries by reuse or importance. Although our mechanism treats all tensors uniformly, the design is pluggable. For instance, a ranker can drive $X$ such that only top-ranked tensors occupy the page-cache path, while the others are routed to the NVMe-direct path.

\vspace{1pt}
\noindent\textbf{Bind \& Materialize.} 
\emph{Bind} maps residency classes to storage. Group~1 KPUs are assigned to filesystem extents (using \texttt{mmap}), and Group~2 KPUs to NVMe namespace LBA extents (detailed in \S\ref{section:design_2}).
A fixed DRAM reservation is brittle under time-varying memory pressure (risking OOM), whereas the OS page-cache via \texttt{mmap} provides best-effort DRAM caching.

To isolate paths, we separate the NVMe space using partitions or distinct namespaces. During inference, each request is dispatched by the binary decision \(x_i\), effectively \emph{materializing} the logical binding. For read and write operations, Group~1 utilizes the page-cache under filesystem control, whereas Group~2 directly accesses reserved LBA blocks via NVMe-direct.

\begin{algorithm}[t]
\SetAlCapFnt{\footnotesize}
\SetAlCapNameFnt{\footnotesize}
\scriptsize
\caption{KPU Residency Planner (parameterized by $X$)}
\label{alg:classify_kpu_to_residency_class}
\KwIn{The set of KPU for all layers \(\{(K^{(i)},V^{(i)})\}_{i=1}^{L}\),\\ 
\quad\quad\quad Size of a single KPU (either K or V) \(S_{kpu}\), knob $X \in [0, B_{pc}]$}
\KwOut{A binary decision \(\{x_i\}_{i=1}^{L}\) where \(x_i\in\{0,1\}\)}
\BlankLine
\(n_1 \gets \min(\lfloor X/(2 \cdot S_{kpu})) \rfloor, L)\)\;
\For{\(i \gets 1\) \KwTo \(L\)}{
\tcp{\scriptsize Residency Class: Group~1, page-cache path}\
  \lIf{\(i \le n_1\)}{ 
  \(x_i \gets 1\);\ \textsc{FileBind}\((K^{(i)})\);\ \textsc{FileBind}\((V^{(i)})\)
}
  \tcp{\scriptsize Residency Class: Group~2, NVMe-direct path}\
  \lElse{
  \(x_i \gets 0\);\ \textsc{LBABind}\((K^{(i)})\);\ \textsc{LBABind}\((V^{(i)})\)
}
}

\Return{\(\{x_i\}\)}
\end{algorithm}

\begin{table}[b]
\vspace{5pt}
\footnotesize
\setlength{\tabcolsep}{4pt}
\centering
\caption{Minimal tensor I/O unit for a single KV component (K or V), FP16/BF16. General formula: $\text{bytes} = B \times H \times D \times 2$.}
\label{tab:min-tensor-unit-kv}
\begin{tabular*}{\linewidth}{@{\extracolsep{\fill}}l c c l l@{}}
\toprule
\textbf{Model} & $H$ & $D$ & \textbf{Formula (bytes)} & \textbf{Size (KiB)} \\
\midrule
OPT-1.3B  & 32 & 64  & $B \times 32 \times 64 \times 2$  & $\mathbf{4}\,\mathrm{KiB} \times B$ \\
OPT-6.7B  & 32 & 128 & $B \times 32 \times 128 \times 2$ & $\mathbf{8}\,\mathrm{KiB} \times B$ \\
OPT-13B   & 40 & 128 & $B \times 40 \times 128 \times 2$ & $\mathbf{10}\,\mathrm{KiB} \times B$ \\
\bottomrule
\end{tabular*}
\end{table}

\begin{figure*}[!t]
	\centering
    \includegraphics[width=0.95 \linewidth]{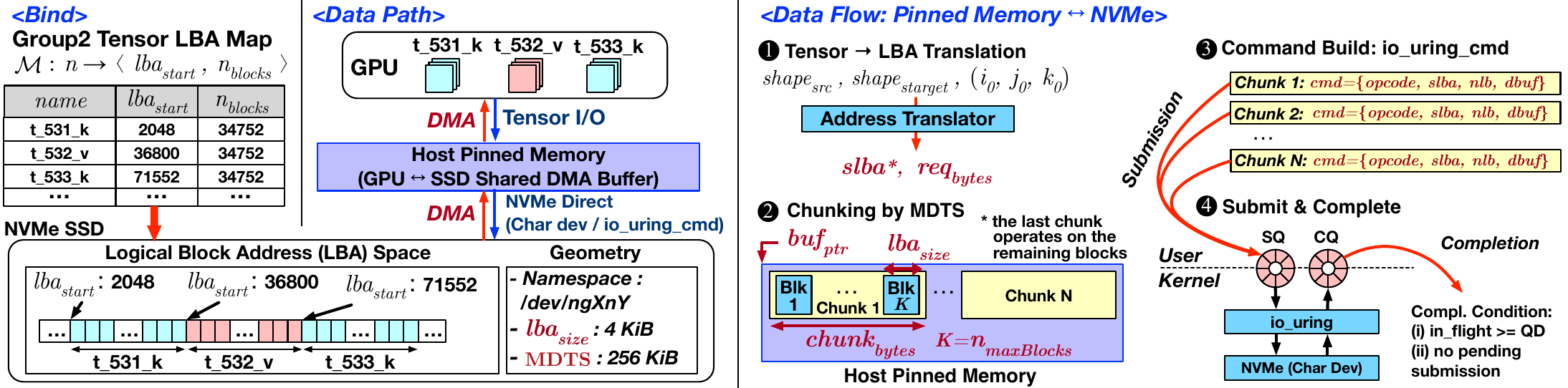}
    \caption{\small Architecture of the NVMe-direct path: Bind, Data Path, and Data Flow.}
    \label{fig:design2}
    \vspace{-15pt}
\end{figure*}

\subsection{NVMe-direct with Sequential-LBA Placement}
\vspace{-2pt}
\label{section:design_2}
The NVMe-direct design accelerates Group~2 KPUs (\S\ref{section:design_1}) by removing kernel storage stack overheads and the tensor--LBA mismatch. The key idea is to bypass the kernel-stack (including the filesystem) and map tensors to a contiguous LBA range in the NVMe namespace.

\vspace{1pt}
\noindent\textbf{Precondition.} 
The tensor size defines the minimum I/O unit and must be an integer multiple of the device LBA size (typically 512\,B or 4\,KiB). The minimum tensor layout is the single-token (\(S{=}1\)) \(K\) or \(V\) tensor with shape \((1, B\!\cdot\!H, D)\), where \(S, B, H, \text{and } D\) denote sequence length, batch size, heads, and per-head dimension. For OPT models~\cite{zhang2022opt}, this alignment generally holds when \(B{\ge}1\) (Table~\ref{tab:min-tensor-unit-kv}). For edge cases (e.g., OPT-13B \(\approx\)10\,KiB, which is not a 4\,KiB multiple), an even $B$ is selected to ensure alignment.

\vspace{1pt}
\noindent\textbf{LBA Bind.}
The \emph{\textless Bind\textgreater} process in Figure~\ref{fig:design2} maps Group~2 KPUs onto a contiguous namespace extent via a hash map \(\mathcal{M}\) from tensor ID \(n\) to its start LBA and block count:
{%
\setlength{\abovedisplayskip}{2pt}%
\setlength{\belowdisplayskip}{2pt}%
\begin{flalign}
& \mathcal{M}:\ n\ \to\ \langle\, lba_{start},\ n_{blocks}\,\rangle && \label{eq:mapping}\\[-2pt]
& \mathrm{extent}(n)\;=\;\big[\, lba_{start}(n),\ lba_{start}(n)+n_{blocks}(n)\,\big) && \label{eq:extent}
\end{flalign}
}
The binding adheres to three invariants: (i) \textit{Alignment} (tensor I/O unit is a multiple of LBA size), (ii) \textit{Disjointness} (extents do not overlap), and (iii) \textit{Contiguity} (each next tensor begins where the previous ends). Only the first tensor’s \(lba_{start}\) is user-specified. Others follow:
{%
\setlength{\abovedisplayskip}{2pt}%
\setlength{\belowdisplayskip}{2pt}%
\begin{flalign}
& n_{blocks}(n) = \mathrm{bytes}(\text{tensor }n)\,/\,lba_{size} && \label{eq:nblocks}\\[-2pt]
& lba_{start}(n{+}1) = lba_{start}(n) + n_{blocks}(n) && \label{eq:lba-recur}
\end{flalign}
}
Here, $\mathrm{bytes}()$ is the byte size of the tensor. For instance, setting \(lba_{start}(t\_531\_k)=2048\) sequentially determines the start LBAs of \(t\_532\_v,\,t\_533\_k,\ldots\). Thus all Group~2 KPUs occupy one large contiguous extent, preserving the KV tensor’s logical sequential stream at the NVMe controller.

\vspace{1pt}
\noindent\textbf{GPU$\leftrightarrow$NVMe Co-DMA over Host Pinned Memory.} 
In the case of the page-cache path (Group~1), write operations follow a 3-hop sequence: (1) GPU$\to$pinned memory (DMA), (2) pinned memory$\to$page-cache (\texttt{memcpy}), and (3) page-cache$\to$NVMe (DMA). Read operations follow the inverse sequence (or 2-hop path on a page-cache hit).

Conversely, the \emph{\textless Data Path\textgreater} in Figure~\ref{fig:design2} illustrates the NVMe-direct path (Group~2) for tensor I/O between the GPU and NVMe.
By bypassing the page-cache, this path removes \texttt{memcpy}  and forms a symmetric 2-hop flow. Specifically, writes follow GPU$\to$buffer$\to$NVMe \texttt{WRITE}, while reads follow NVMe \texttt{READ}$\to$buffer$\to$GPU.
We use a Co-DMA scheme, where the GPU and NVMe share the same per-thread pinned buffer, which is already allocated for GPU DMA.

Three factors validate this design:
(i) \textit{DMA buffer} constraints are aligned, meaning buffers are page-locked and 4\,KiB-aligned, allowing both the GPU (via IOMMU/IOVA) and NVMe (via PRP/SGL) to access the same buffer. 
(ii) \textit{Exclusive ownership} ensures copy-threads use disjoint pinned regions and operate unidirectionally (read or write) per tensor.
(iii) \textit{Dual views} enable the same memory to serve as a structured tensor to the GPU and a byte array to NVMe.

\begin{algorithm}[t]
\SetAlCapFnt{\footnotesize}
\SetAlCapNameFnt{\footnotesize}
\scriptsize
\caption{Tensor-Index–to–LBA Translation}
\label{alg:tensor_index_to_lba}
\DontPrintSemicolon
\KwIn{$name$ \textbf{(tensor id)}, $\textit{shape}_{src}{=}(f_0,f_1,f_2)$ \textbf{(source tensor shape)}, $\textit{shape}_{tgt}{=}(d_0,d_1,d_2)$ \textbf{(target tensor shape)}, $(i_0,j_0,k_0)$ \textbf{(offset indices in target)}, $e$ \textbf{(bytes per element, e.g., 2 for FP16/BF16)}, $lba_{size}$ \textbf{(bytes)}}
\KwOut{$slba^{\star}$, $req_{bytes}$}
\BlankLine
\textbf{Precondition (as stated in the section).} The minimal tensor I/O unit is an integer multiple of $lba_{size}$ and placement starts on unit boundaries. Thus $req_{bytes}$ and $offset_{bytes}$ are multiples of $lba_{size}$.\;
\BlankLine
\tcp*[l]{\scriptsize lookup from the tensor$\to$LBA map}
$(lba_{start},\_) \gets \mathcal{M}[name]$\;
\tcp*[l]{\scriptsize row-major element offset and byte offset}
$offset_{elem} \gets ((i_0\cdot d_1 + j_0)\cdot d_2 + k_0)$\;
$offset_{bytes} \gets offset_{elem}\cdot e$\;
\tcp*[l]{\scriptsize anchor LBA and requested payload size}
$slba^{\star} \gets lba_{start} + \big(offset_{bytes}/lba_{size}\big)$\;
$req_{bytes} \gets (f_0\cdot f_1\cdot f_2)\cdot e$\;
\BlankLine
\Return $(slba^{\star},\ req_{bytes})$
\end{algorithm}

\vspace{1pt}
\noindent\textbf{Data Flow: Pinned Memory $\leftrightarrow$ NVMe.}
To execute direct reads or writes on NVMe via \texttt{io\_uring\_cmd}, the high-level tensor I/O request must be translated into device-specific native NVMe I/O parameters.
In the \emph{\textless Data Flow\textgreater} of Figure~\ref{fig:design2}, a copy thread translates tensor info into the target read/write location (LBA) and request size (\bcircled{1}).
Algorithm~\ref{alg:tensor_index_to_lba} looks up \(lba_{start}\) in \(\mathcal{M}\) (line~2), computes the row-major byte offset from \(\mathrm{shape}_{target}\) and indices \((i_0,j_0,k_0)\) (lines~3--4), then derives \(slba^{\star}\) and \(req_{bytes}\) from \(\mathrm{shape}_{source}\) (lines~5--6).

Next, the copy-thread chunks $req_{bytes}$ to fit within the NVMe controller’s \emph{maximum data transfer size} (MDTS), the per-command payload cap (e.g., 256~KB) (\bcircled{2}). 
To maximize throughput while preserving $lba_{size}$ alignment, the chunk size is set to the largest multiple of $lba_{size}$ fitting within the MDTS. 
Consequently, the number of chunks and maximum blocks per chunk are derived as:
{%
\setlength{\abovedisplayskip}{2pt}%
\setlength{\belowdisplayskip}{2pt}%
\begin{flalign} & chunk_{bytes} = \operatorname{align\_down}(\mathrm{MDTS},\,lba_{size}) && \label{eq:chunkbytes}\\[-2pt] & n_{chunks} = \left\lceil \frac{req_{bytes}}{chunk_{bytes}} \right\rceil,\; n_{maxBlocks} = \frac{chunk_{bytes}}{lba_{size}} && \label{eq:nchunks-nmax} 
\end{flalign}
}

For each chunk $N\in[1,n_{chunks}]$, the copy-thread constructs an NVMe native command (\bcircled{3}) with: 
(i) \(\textit{opcode}\in\{\texttt{READ},\texttt{WRITE}\}\),
(ii) $(nsid, slba, nlb)$ denoting the namespace ID, the starting LBA, and the logical block count (where $nlb$ is 0-based, transferring $nlb+1$ blocks),
and (iii) the host data buffer $dbuf$ pointing to the pinned-memory address for chunk $N$.
The per-chunk parameters are:
{%
\setlength{\abovedisplayskip}{2pt}%
\setlength{\belowdisplayskip}{2pt}%
\begin{flalign}
& slba = slba^{\star} + (N-1)\times n_{maxBlocks} && \label{eq:slba}\\[-2pt]
& nlb = \min\!\big(n_{maxBlocks},\,n_{remains}\big) - 1 && \label{eq:nlb}\\[-2pt]
& dbuf = buf_{ptr} + (N-1) \times chunk_{bytes} && \label{eq:dbuf}
\end{flalign}
}
Per-chunk NVMe commands are submitted asynchronously to the submission queue (SQ), and completions arrive on the completion queue (CQ) (\bcircled{4}). 
The copy-thread harvests completion queue entries (CQEs) when in-flight requests reach the queue depth (QD) or when draining remaining chunks. 
A tensor transfer request, whether read or write, is finalized once all its constituent chunk CQEs are collected.

\vspace{1pt}
\noindent\textbf{Dataset Management--Deallocate (TRIM).} 
On context teardown, allocated LBA space is reclaimed via NVMe \emph{Dataset Management (DSM)} \emph{deallocate}. 
For each tensor, we query \(\mathcal{M}\) for \((lba_{start}, n_{blocks})\) and submit TRIM via \texttt{io\_uring\_cmd}.
This command explicitly hints to the NVMe controller that the specified LBA range no longer contains valid data, enabling efficient and rapid deallocation of large, contiguous KV ranges.

Although TRIM can trigger internal garbage collection (GC) potentially affecting latency, since concurrent requests are processed as batched tensors or scheduled serially in a single-GPU system, our architecture effectively eliminates LBA fragmentation.
Consequently, the device maintains a near-ideal Write Amplification Factor (WAF) close to 1, rendering GC overhead negligible.

\begin{figure}[t]
	\centering
    \includegraphics[width=0.96 \linewidth]{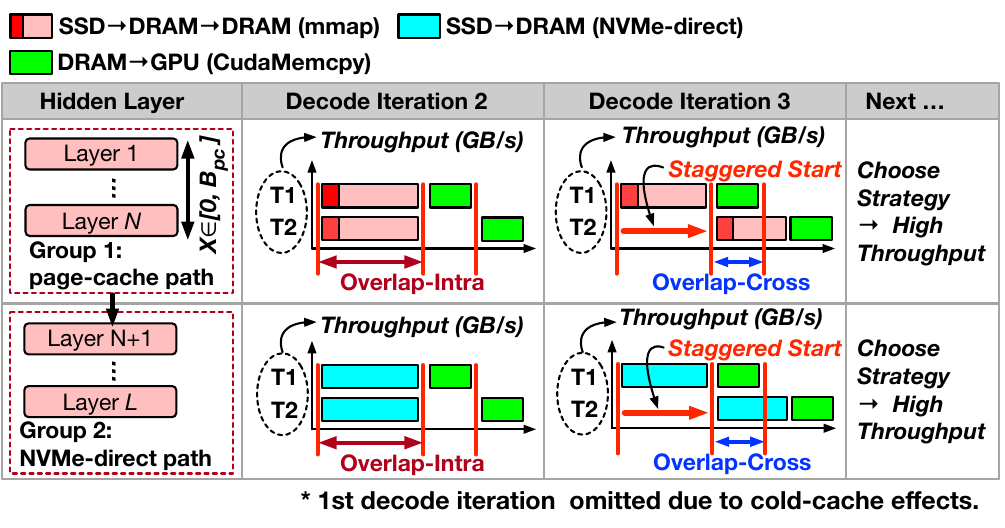}
    \vspace{-3pt}
    \caption{\small Adaptive pipeline parallelism with two overlap strategies.}
    \label{fig:design_3}
    \vspace{-3pt}
\end{figure}

\subsection{Pipeline Parallelism between GPU DMA and Storage I/O}
\vspace{-2pt}
\label{section:design_3}
Building on the dual-path and NVMe-direct design,
this section presents a pipeline parallelism to maximize
throughput by overlapping GPU DMA and storage I/O in a multi copy-thread
environment. This dynamically selects between two pipeline strategies,
\textit{overlap-intra} and \textit{overlap-cross}.

\vspace{1pt}
\noindent{\textbf{Parallelism Types and Constraints.}}
In a multi copy-thread setting, two threads can process the \textit{K} and
\textit{V} tensors of each layer concurrently (e.g., thread~1 for \textit{K}
and thread~2 for \textit{V}). We categorize the parallelism into
two types. (i) \textit{Overlap-Intra} overlaps I/O operations within the same
hardware path (e.g., parallel NVMe reads), which improves bandwidth
utilization but may introduce contention when the storage is saturated.
While NVMe SSDs support parallel I/O via multiple hardware queues,
copy streams issued by multiple threads are effectively serialized on the GPU~\cite{cuda-programming-guide}.
(ii) \textit{Overlap-Cross} overlaps I/O across
different hardware paths (e.g., NVMe reads and GPU DMA), enabling effective
latency hiding by utilizing independent hardware and PCIe resources.

\vspace{1pt}
\noindent{\textbf{Adaptive Pipeline Strategy Selection.}}
Because GPU DMA and storage I/O naturally overlap during the prefill phase,
we focus on the decode phase. Figure~\ref{fig:design_3} illustrates the comparison and selection of two pipeline strategies for each path (Group~1 and Group~2) during decode phase.

\textit{(1) Warm-up:}
To exclude transient latency caused by cold page-cache effects in the Group~1 path, 
the first decode iteration is omitted from measurement.
\textit{(2) Iteration~2 (Overlap-Intra):}
Storage reads from two copy-threads are parallelized,
while GPU H2D DMA is executed serially.
This maximizes storage bandwidth utilization when the storage is not saturated.
\textit{(3) Iteration~3 (Overlap-Cross):}
A staggered start delays the storage read on thread~T2,
overlapping it with GPU DMA issued by thread~T1 on a different
hardware path.
Under saturated storage bandwidth, this mitigates contention and improves
overall pipeline throughput.
\textit{(4) Strategy Selection:}
The throughput of the two iterations is compared for each group,
and the higher-performing strategy is fixed for subsequent decoding iterations.
The adaptive pipeline incurs only a bounded overhead from lightweight throughput monitoring and at most one trial iteration for strategy comparison. This constant cost is amortized as generation length increases.

\begin{table}[b]
\vspace{5pt}
\footnotesize
\setlength{\tabcolsep}{4pt}
\renewcommand{\arraystretch}{1.22}

\centering
\caption{Evaluation configurations for KV-cache offloading}
\vspace{-4pt}
\label{tab:cfg-variants}
\begin{tabularx}{\linewidth}{@{}l >{\raggedright\arraybackslash}X@{}}
\toprule
\textbf{Label} & \textbf{Description} \\
\midrule
Baseline &
All KV accesses use the page-cache path. \\
CachePolicy-Only &
$X=B_{pc}$ (upper bound). Page-cache path only. Group~1 remains resident in the page-cache. Group~2 remains on the page-cache path, not the NVMe-direct path, and is
proactively evicted via \texttt{posix\_fadvise(POSIX\_FADV\_DONTNEED)}. \\
NVMe-direct-Only &
$X=0$ (lower bound). All KV accesses use the NVMe-direct path via \texttt{io\_uring\_cmd}, the page-cache is unused. \\
\dualblade{} &
$X=B_{pc}$ (upper bound). Dual-path design: tensors up to $B_{pc}$ use the page-cache path (Group~1) and the remainder uses the NVMe-direct path (Group~2). \\
\bottomrule
\end{tabularx}
\end{table}

\section{Evaluation}
\label{sec:evaluation}
\vspace{-2pt}
\subsection{Experimental Setup}
\vspace{-2pt}
\label{sec:expr_setup}

\noindent\textbf{Experimental Setup.}
All experiments were conducted on a single-GPU system equipped with an
Intel Core Ultra 7 265K processor (20~cores at 3.8\,GHz), 16\,GB of host memory,
and an NVIDIA RTX 5060~Ti (16\,GB). The software environment includes Ubuntu~24.04.2, Linux~6.8.0, liburing~2.5, and PyTorch~2.8.0 (+CUDA~12.8).
We employed FlexLLMGen~\cite{sheng2023flexgen} (commit \texttt{004ffef}) as the experimental framework.

To assess device-dependent effects and investigate how storage media characteristics impact end-to-end inference latency, we evaluate two NVMe SSDs with distinct classes and I/O constraints: SSD~A (Samsung PM9D3a, PCIe Gen5, 4\,KiB LBA, 256\,KiB MDTS) and SSD~B (Samsung 990 PRO, PCIe Gen4, 512\,B LBA, 2\,MiB MDTS).
This comparison validates whether our architectural benefits persist across different hardware generations and internal controller specifications, as LBA size and MDTS directly constrain NVMe-direct alignment and per-command transfer granularity.

\vspace{1pt}
\noindent\textbf{Configuration.}
We evaluated the OPT-6.7B model~\cite{facebook_opt6.7b}. In FlexLLMGen, only the KV-cache is offloaded, while parameter offloading is disabled. Unless otherwise noted, the prompt length is 512 tokens, the generation length is 32 tokens, and the batch size is 32. Although we evaluated various parameter settings, the trends remain consistent. Therefore, we report this representative case. 
We use two copy-threads, and additional threads provide no benefit and typically remain idle.

\vspace{1pt}
\noindent\textbf{Comparison.}
We evaluate the effects of three design choices.
CachePolicy-Only isolates the impact of the \emph{KV Residency Policy}.
NVMe-direct-Only evaluates the benefit of the \emph{NVMe-direct} I/O path.
\dualblade{} combines both designs as a dual-path NVMe-direct approach.
Baseline corresponds to vanilla FlexLLMGen.
The detailed configurations are summarized in Table~\ref{tab:cfg-variants}.
The third key component, \emph{Pipeline Parallelism}, is applied in a
multi copy-thread environment to CachePolicy-Only, NVMe-direct-Only,
and \dualblade{}.

\begin{figure}
    \centering 
    \includegraphics[width=0.96\linewidth]{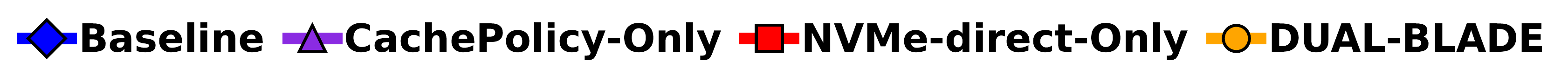}
    \vspace{-2pt} 
    \\[0pt]

    \subfloat[\small Prefill (SSD~A) ]{%
    \vspace{-3pt}
    \includegraphics[width=0.48\linewidth]{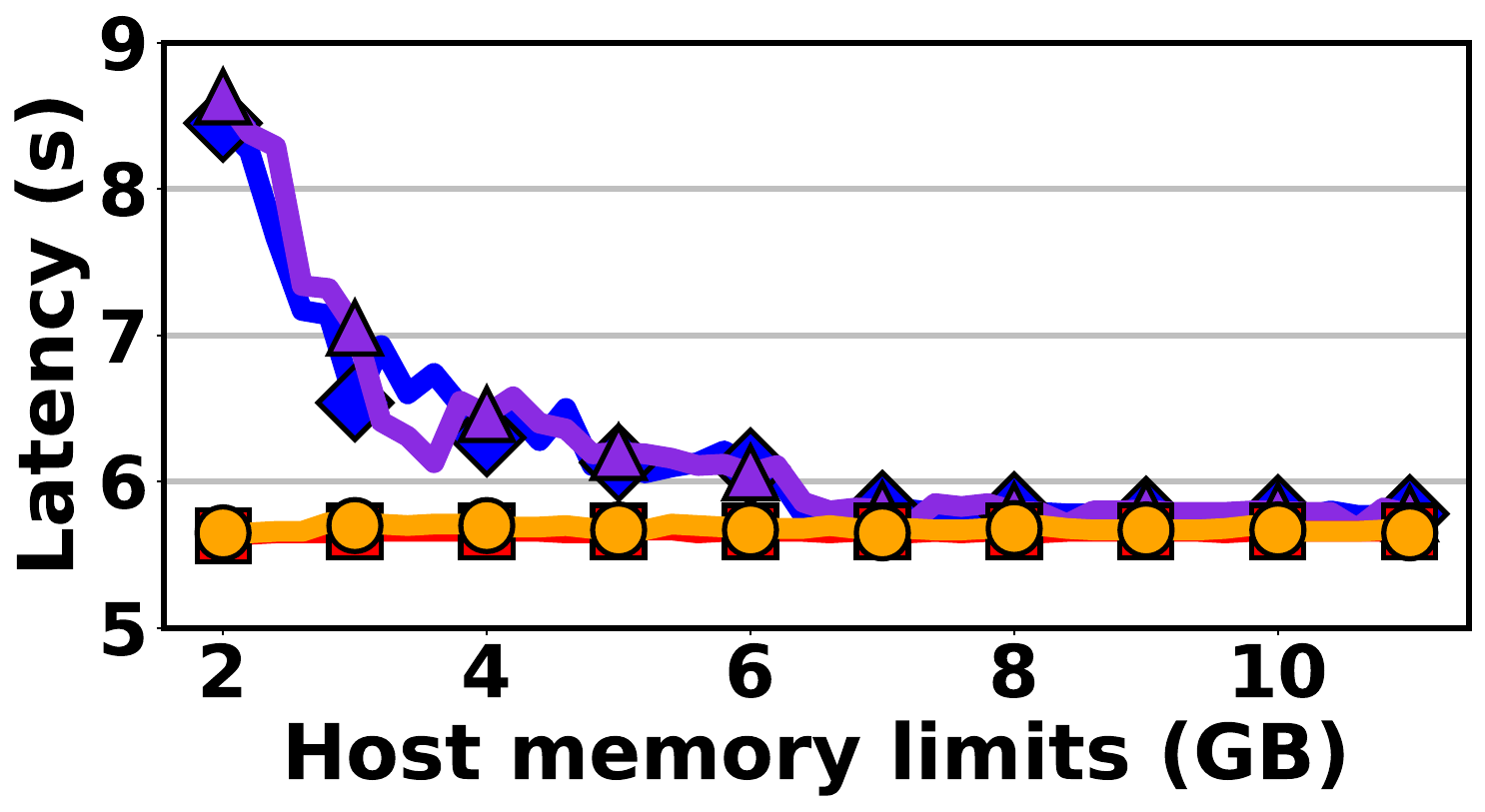}}
    \vspace{2pt}
    \hfil
    \subfloat[\small Decode (SSD~A)]{%
    \vspace{-3pt}
    \includegraphics[width=0.48\linewidth]{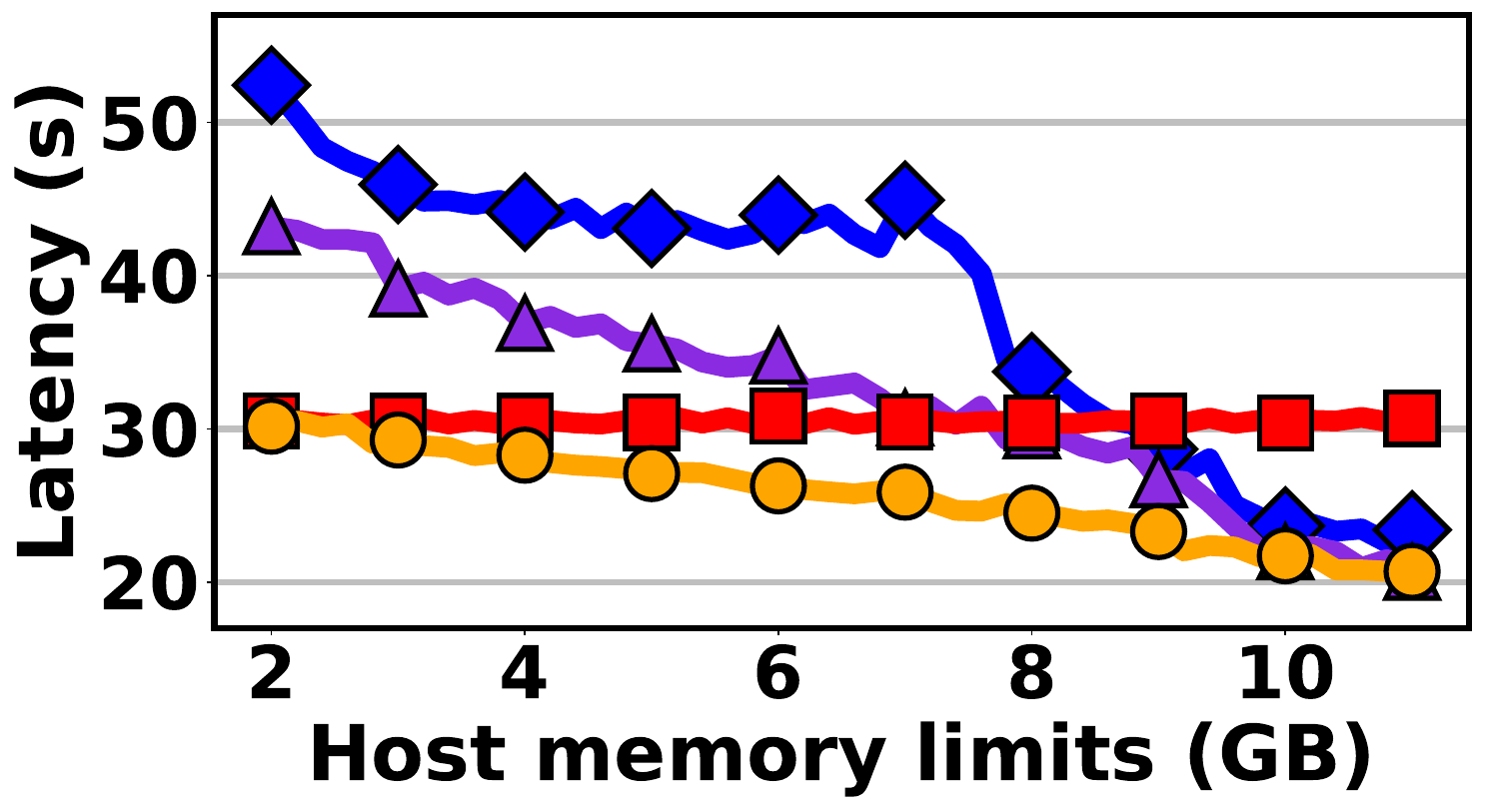}}
    \vspace{2pt}
    \hfil
    \subfloat[\small Prefill (SSD~B)]{%
    \vspace{-3pt}
    \includegraphics[width=0.48\linewidth]{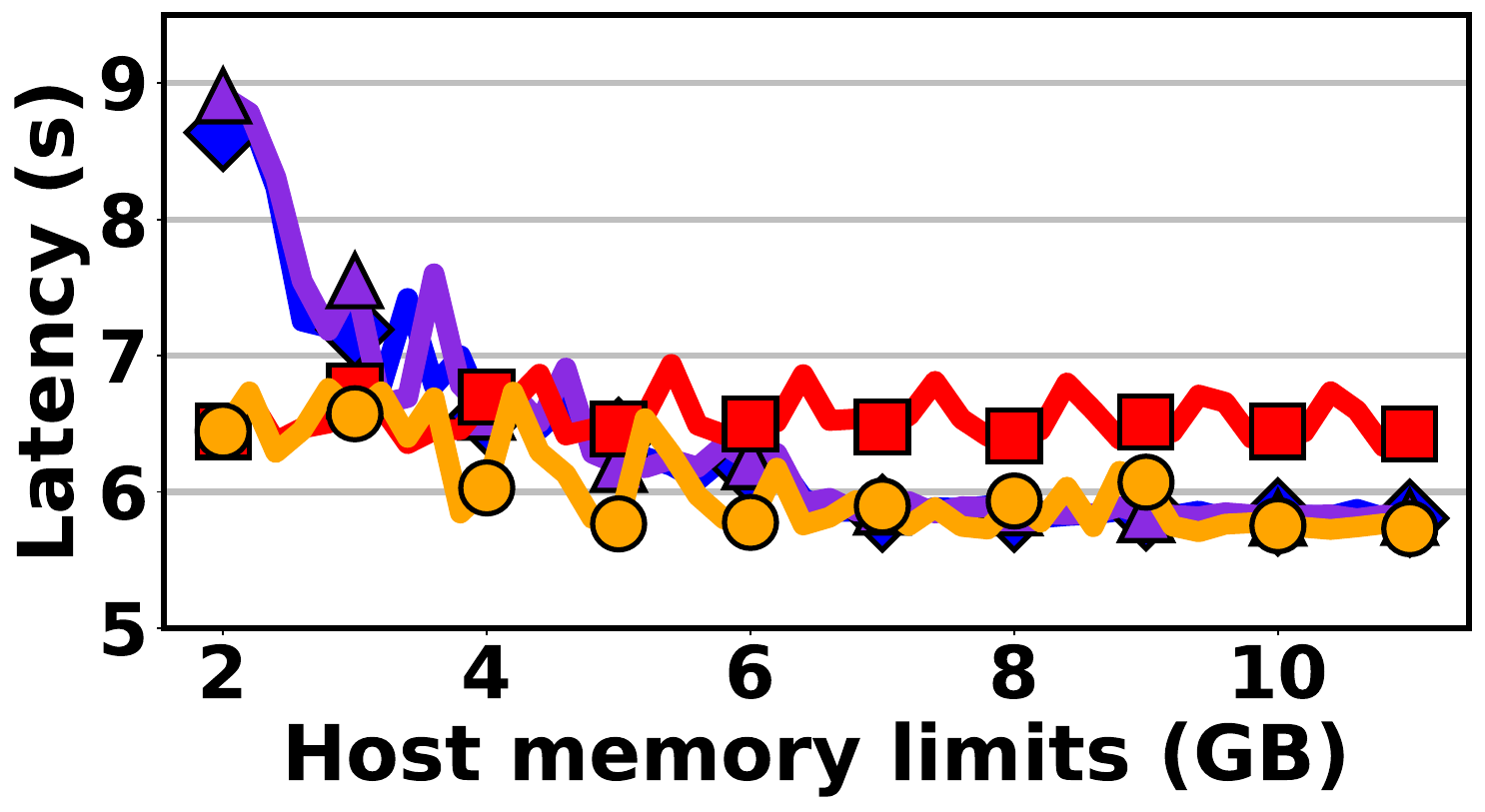}}
    \hfil
    \subfloat[\small Decode (SSD~B)]{%
    \vspace{-3pt}
    \includegraphics[width=0.48\linewidth]{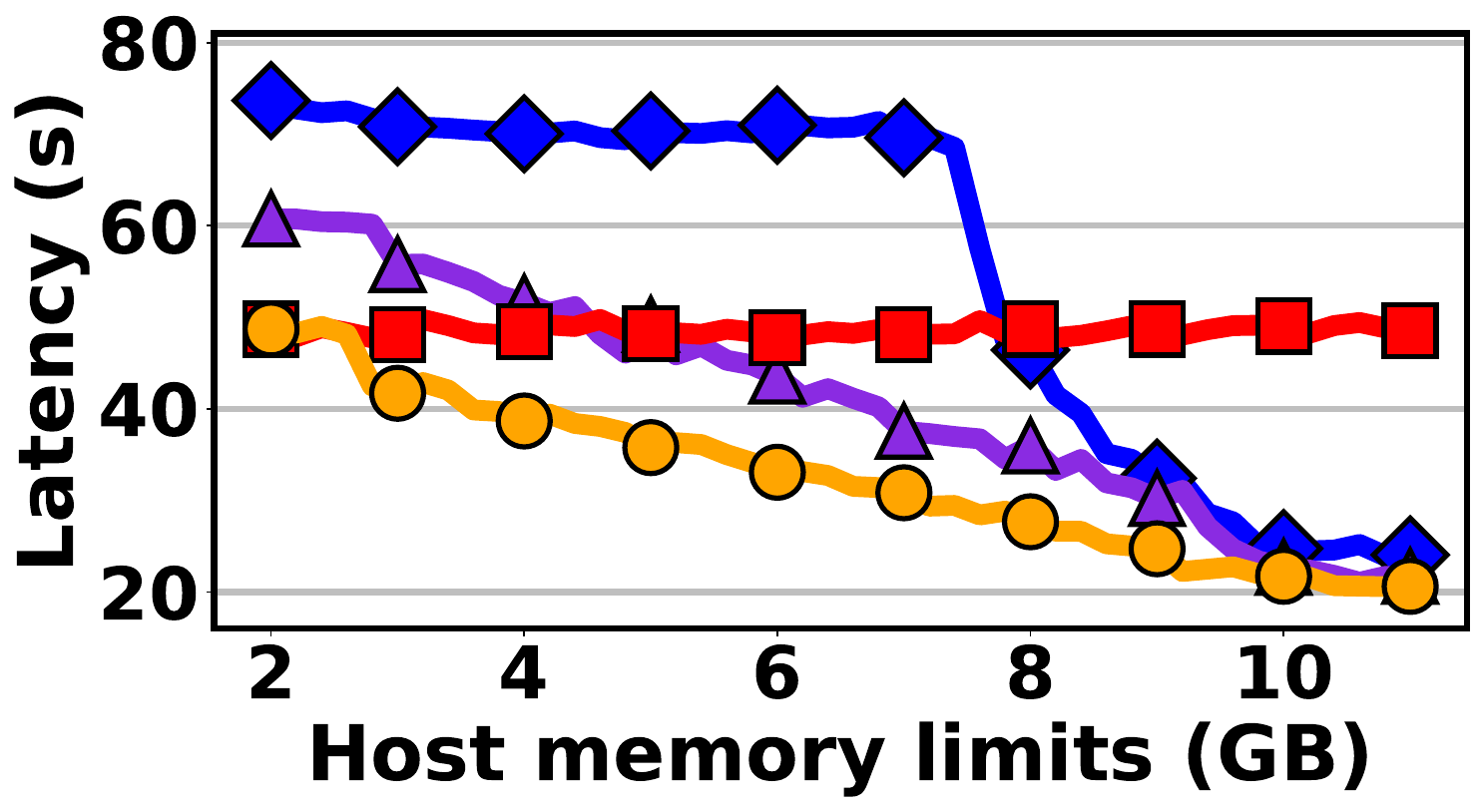}}
    \hfil
    \vspace{-3pt}
    \caption{\small End-to-end inference latency of OPT-6.7B on SSD A and SSD B under varying host memory limits (lower is better).}
    \vspace{-3pt}
\label{fig:eval_serving_lat}
\end{figure}

\subsection{LLM Serving Evaluation: Prefill and Decode}
\vspace{-2pt}
\label{sec:serv_eval}
To evaluate end-to-end inference performance, we measured prefill and decode latencies across all comparison groups.
We executed the workload repeatedly under varying host memory limits (2--11\,GB) to capture page-cache dynamics.
In these experiments, we employed a multi copy-thread configuration and evaluated two distinct SSDs, SSD~A and SSD~B, to verify performance consistency, as shown in Figure~\ref{fig:eval_serving_lat}.

We observe consistent trends across the results.
First, CachePolicy-Only effectively avoids the page-cache thrashing observed in the Baseline during the decode phase by employing the KV residency policy. Consequently, it exhibits a linear reduction in latency proportional to the host memory limit.

Second, NVMe-direct-Only, which bypasses the page-cache, exhibits constant latency regardless of host memory limits, as expected.
In decode, while this stability is beneficial under memory pressure, it fails to leverage DRAM as memory limits increase, resulting in the lowest performance.

Finally, \dualblade{} consistently demonstrates superior performance across the full range of host-memory limits.
The results confirm that it mitigates write stalls during prefill (specifically in the 2--7\,GB range) and eliminates page-cache thrashing during decode.
This robustness stems from the adaptive dual-path KV residency, which partitions tensors based on the available page-cache budget, routing excess tensors to the NVMe-direct path (Group~2) under tight memory constraints while increasingly leveraging the page-cache path (Group~1) as the host-memory limit increases.
We next summarize the quantitative gains for each SSD.
Across the 2--11\,GB memory range, on SSD~A, \dualblade{} reduces prefill latency by up to 33.1\% and decode latency by 8.2--42.4\%.
On SSD~B, it achieves prefill reductions of up to 25.4\% and decode reductions of 11.7--57.8\%.

Our results demonstrate that \dualblade{} consistently reduces latency across different storage tiers. This confirms that the standard I/O stack creates a bottleneck regardless of the underlying hardware, highlighting the universality of our solution.
Next, we provide a detailed evaluation quantifying the effectiveness of each individual design component.

\subsection{Mitigating Page-Cache Thrashing}
\vspace{-2pt}
\label{sec:avoid_page_cache}
\begin{wrapfigure}{l}{0.44\linewidth}
    \vspace{-10pt}
    \includegraphics[width=\linewidth]{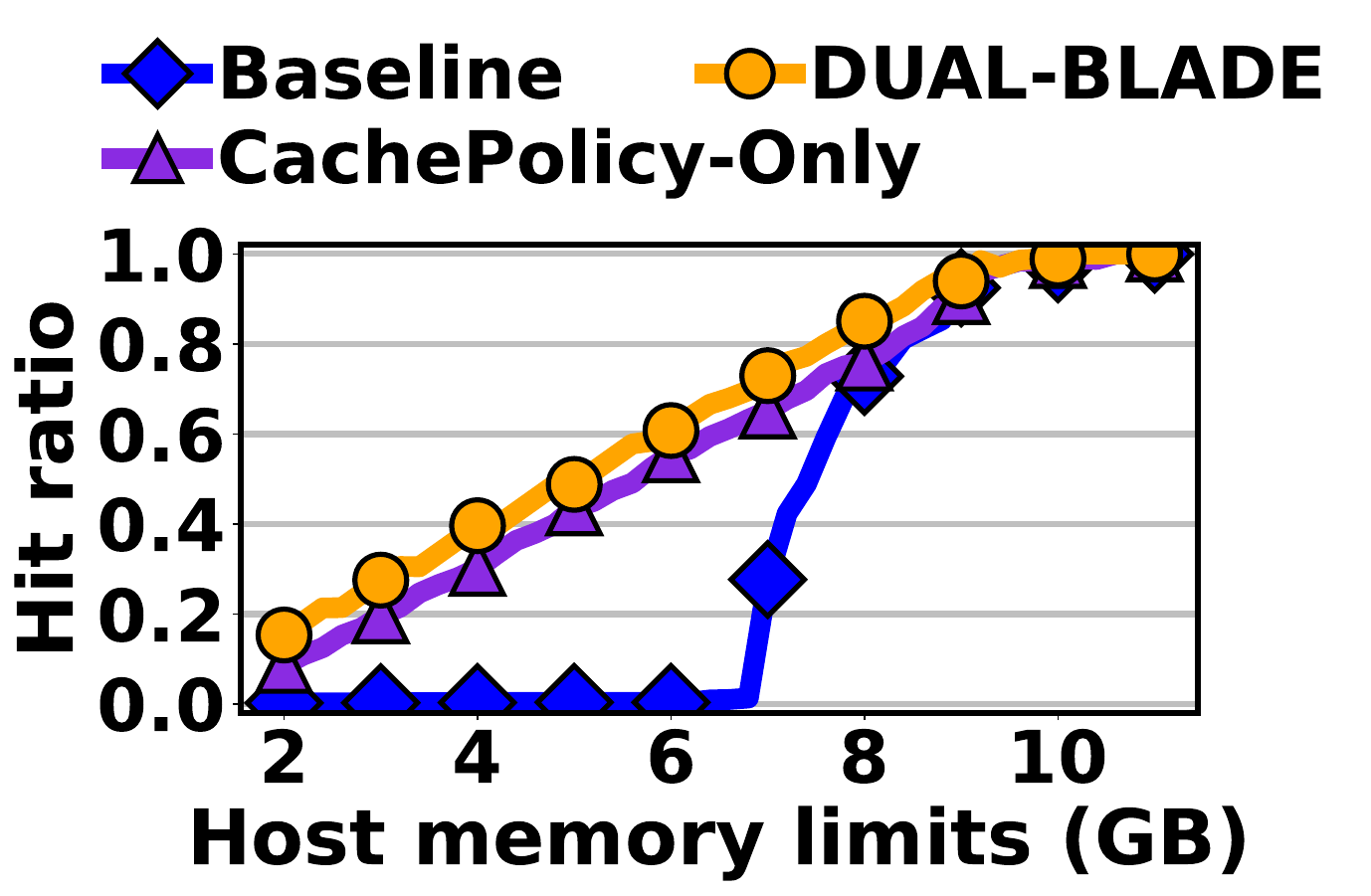}
    \caption{\small Page-cache hit ratio.}
    \vspace{-10pt}
    \label{fig:eval_hit_ratio}
\end{wrapfigure}

Dual-path KV residency (§\ref{section:design_1}) allocates KV tensors to Group~1 and Group~2 to reduce page-cache thrashing and improve the hit ratio in the decode phase.

Figure~\ref{fig:eval_hit_ratio} shows page-cache hit ratios across host-memory limits, defined as the fraction of total read bytes served from the page-cache.
As noted in §\ref{sec:motiv_1}, Baseline exhibits a thrashing zone at 2--7\,GB. By contrast, both CachePolicy-Only and \dualblade{} show a linear increase in hit ratios, effectively mitigating thrashing. 
However, \dualblade{} achieves superior hit ratios due to complete path separation. 
In contrast, CachePolicy-Only relies on \texttt{posix\_fadvise} (to proactively evict data), which inevitably incurs overhead as data must traverse the page-cache.

Furthermore, \dualblade{} resolves the page-cache write stall issue during the prefill phase (§\ref{sec:motiv_1}). Conversely, CachePolicy-Only remains susceptible to these stalls, as it cannot avoid the initial write to the page-cache.

\begin{table}[!b]
  \vspace{5pt}
  \centering
  \begin{threeparttable}
    \scriptsize                       
    \setlength{\tabcolsep}{2pt}       
    \renewcommand{\arraystretch}{1.05}
    \caption{\small Per-tensor I/O latency (between pinned memory and NVMe) and device utilization}
    \label{tab:per-tensor-io}
    \vspace{-4pt}
    \begin{tabular}{@{}l rr rr rr rrr@{}} 
      \toprule
      & \multicolumn{6}{c}{Per-tensor I/O latency (ms)} & \multicolumn{3}{c}{NVMe busy (\%)} \\
      \cmidrule(lr){2-7}\cmidrule(l){8-10}
      \multirow{2}{*}{Config} &
        \multicolumn{2}{c}{Prefill W.} &
        \multicolumn{2}{c}{Decode R.} &
        \multicolumn{2}{c}{Decode W.} &
        \multirow{2}{*}{\makecell{Prefill \\ W.}} &
        \multirow{2}{*}{\makecell{Decode \\ R.}} &
        \multirow{2}{*}{\makecell{Decode \\ W.}} \\
      & avg. & 99th & avg. & 99th & avg. & 99th & & & \\
      \midrule
      \multicolumn{10}{l}{\textit{SSD A}} \\
      Baseline    & 35.79 & 109.40 & 18.80 & 79.59 & 4.91 & 8.26 & 44.79 & 54.58 & 100.00 \\
      \dualblade{}        & \textcolor{blue}{\textbf{21.18}} & \textcolor{blue}{\textbf{21.26}}  & \textcolor{blue}{\textbf{11.17}}  & \textcolor{blue}{\textbf{12.67}}  & \textcolor{blue}{\textbf{0.06}} & \textcolor{blue}{\textbf{0.06}} & \textcolor{blue}{\textbf{100.00}}& \textcolor{blue}{\textbf{98.45}} & 100.00 \\
      
      \addlinespace[3pt]\cmidrule(lr){1-10}
      \multicolumn{10}{l}{\textit{SSD B}} \\
      Baseline    & 41.15 & 148.95 & 32.29 & 85.38 & 4.47 & 8.98 & 89.15 & 86.99 & 100.00 \\
      \dualblade{}        & \textcolor{blue}{\textbf{22.85}} & \textcolor{blue}{\textbf{94.00}}  & \textcolor{blue}{\textbf{19.03}}  & \textcolor{blue}{\textbf{23.64}}  & \textcolor{blue}{\textbf{0.06}} & \textcolor{blue}{\textbf{0.15}} & \textcolor{blue}{\textbf{100.00}} & \textcolor{blue}{\textbf{94.58}} & 100.00 \\
      \bottomrule
    \end{tabular}

    \begin{tablenotes}[flushleft]
      \footnotesize
      \item \textit{Note.} 
  NVMe busy reports the device-level utilization
over the duration of the corresponding tensor I/O (per-tensor, not job-wide average). Per-tensor I/O sizes include a 128~MB (prefill write), 128--135~MB (decode read),
and 256~KB (decode write).
    \end{tablenotes}
  \end{threeparttable}
\end{table}

\subsection{Maximizing NVMe Utilization and Throughput}
\vspace{-2pt}
\label{sec:elevating_nvme_busy}
NVMe-direct (\S\ref{section:design_2}) shortens the host-to-device I/O path between pinned memory and the NVMe device, increasing device utilization (NVMe busy).
Table~\ref{tab:per-tensor-io} compares per-tensor latency and NVMe busy (\%) for Baseline vs.\ \dualblade{} under a 2\,GB host memory limit, excluding page-cache hits for both configurations to isolate storage I/O effects.

Across SSD~A/B, \dualblade{} lowers mean/99th latency and raises utilization (e.g., 2.2$\times$ prefill-write on SSD~A), saturating device bandwidth.
The efficiency gap is most pronounced for the 256\,KB decode-write, where each tensor write maps to a single NVMe command and thus saturates the device (100\% NVMe busy) in both configurations.
Despite identical device busy, \dualblade{} reduces latency by over 98\% on both SSDs (e.g., 4.91$\to$0.06\,ms on SSD~A).
This indicates that Baseline's latency is dominated by software overheads in the conventional I/O path rather than device processing.

\begin{figure}[t]
    
    \centering 
    \includegraphics[width=0.40\linewidth]{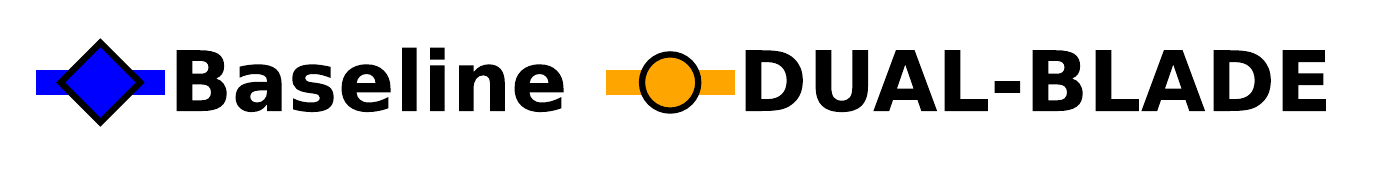}
    \vspace{-2pt} 
    \\[0pt]

    \subfloat[\small Prefill Write (SSD A)]{%
    \vspace{-3pt}
    \includegraphics[width=0.48\linewidth]{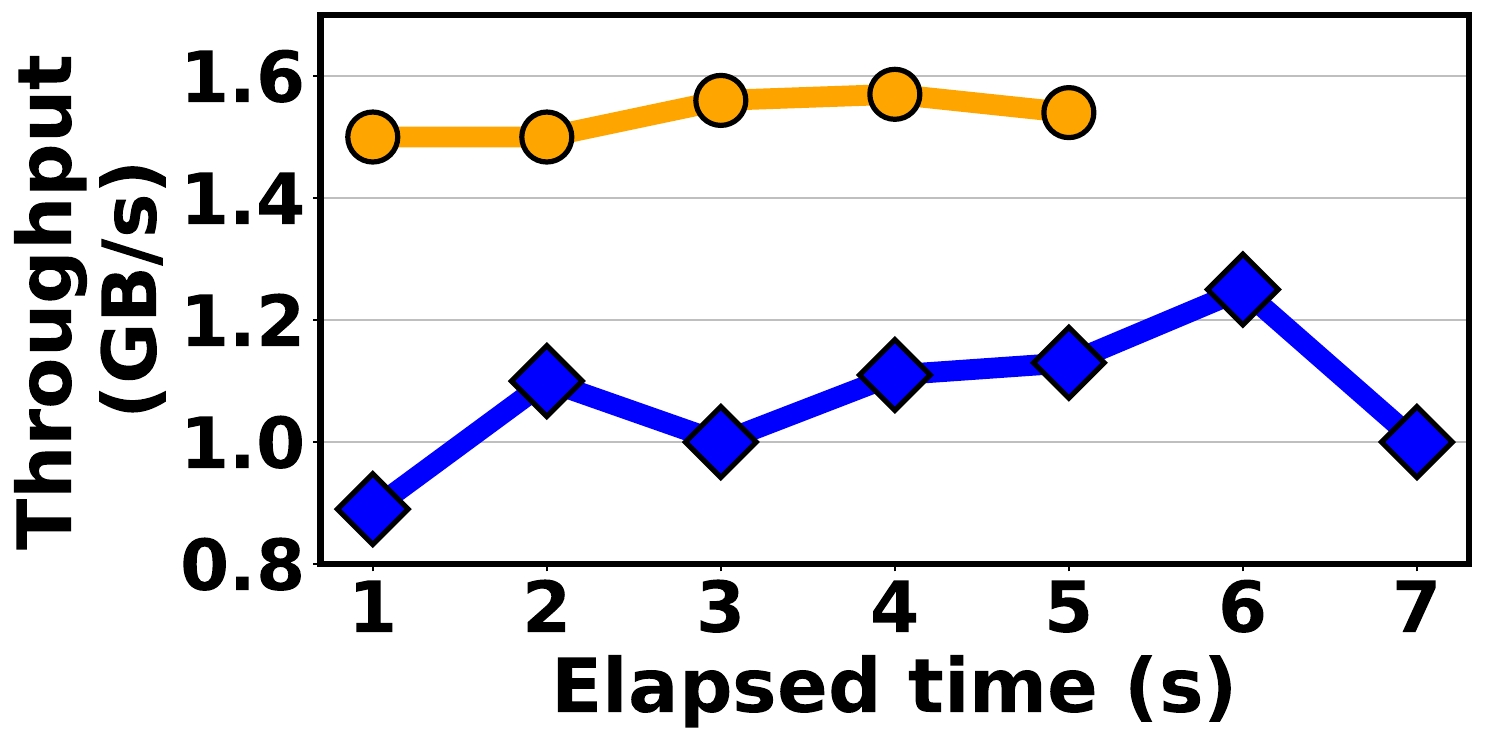}}
    \hfil
    \subfloat[\small Decode Read (SSD A)]{%
    \vspace{-3pt}
    \includegraphics[width=0.48\linewidth]{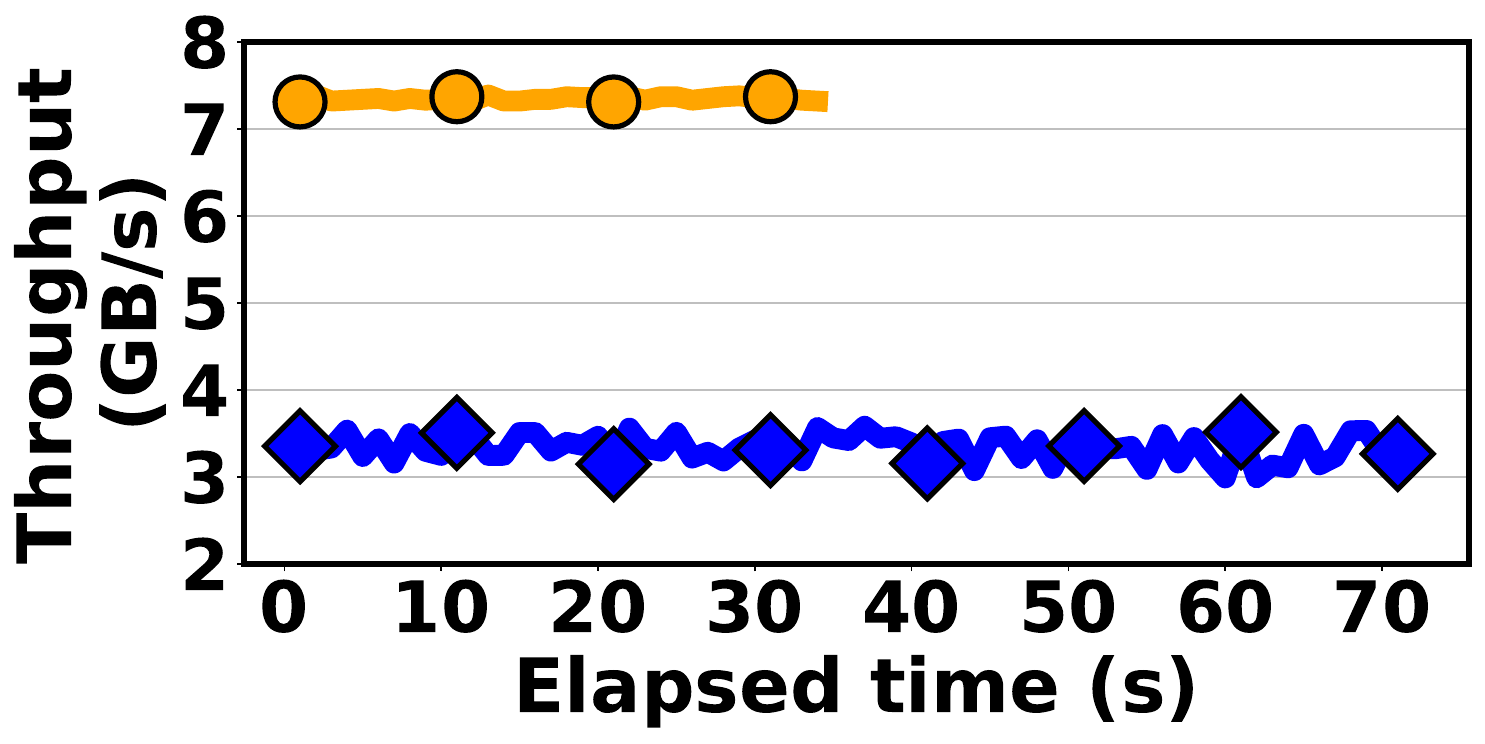}}
    \hfil

    \subfloat[\small Prefill Write (SSD B) ]{%
    \vspace{-3pt}
    \includegraphics[width=0.48\linewidth]{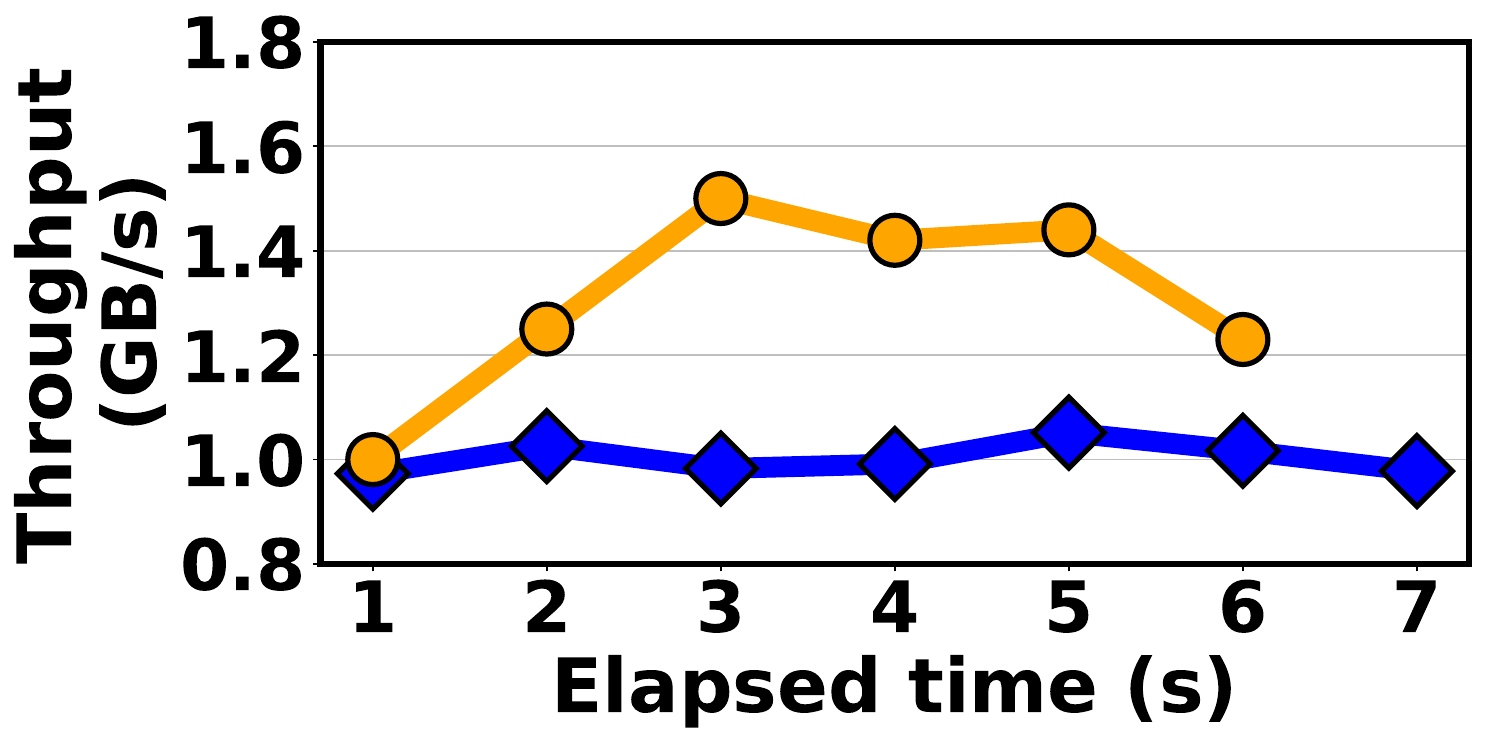}}
    \hfil
    \subfloat[\small Decode Read (SSD B)]{%
    \vspace{-3pt}
    \includegraphics[width=0.48\linewidth]{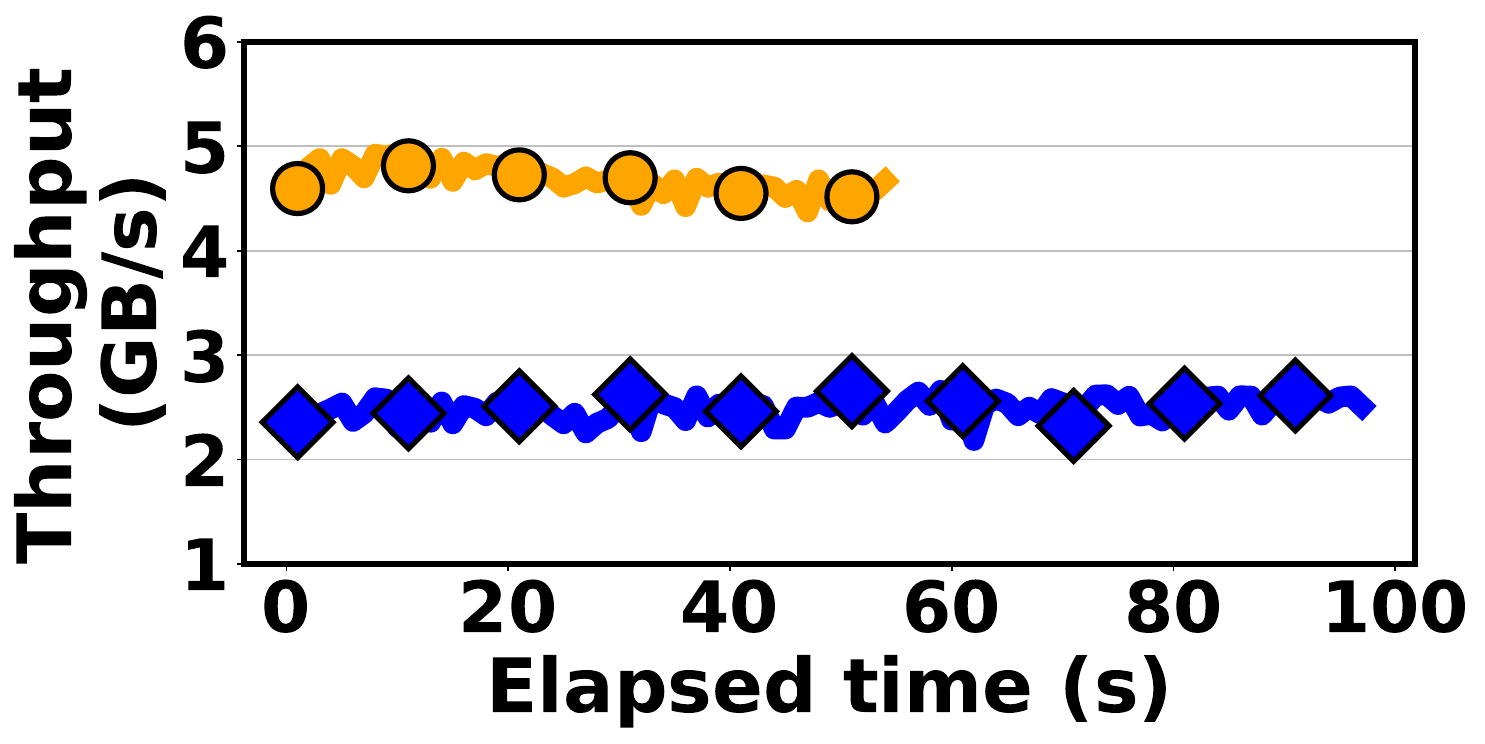}}
    \hfil
    
    \vspace{-3pt}
    \caption{\small Disk throughput (GB/s) over time on SSD~A/B under a 2\,GB memory limit (higher is better).}
	\vspace{-3pt}
\label{fig:ssd_throughput}
\end{figure}

Figure~\ref{fig:ssd_throughput} plots disk throughput over time for prefill-write and decode-read on SSD~A and SSD~B (decode-write omitted for brevity). These results provide direct evidence that the higher NVMe utilization (Table~\ref{tab:per-tensor-io}) of the NVMe-direct path translates into higher realized disk throughput.
Specifically, on SSD A (average throughput), NVMe-direct-Only outperforms Baseline by 43\% in prefill-write and by 120\% (2.20$\times$) in decode-read. 
On SSD B, gains 43\% in prefill-write  and 88\% (1.88×) in decode-read over Baseline.

Collectively, these results confirm that NVMe-direct shortens the conventional kernel storage I/O stack and increases NVMe utilization, which consequently translates into improved overall disk throughput.

\subsection{Sequential LBA Access, Tighter Submit--Complete Latency}
\vspace{-2pt}
\label{sec:flat_lba}

\begin{figure}[t]
    
    \centering 
    \includegraphics[width=0.5\linewidth]{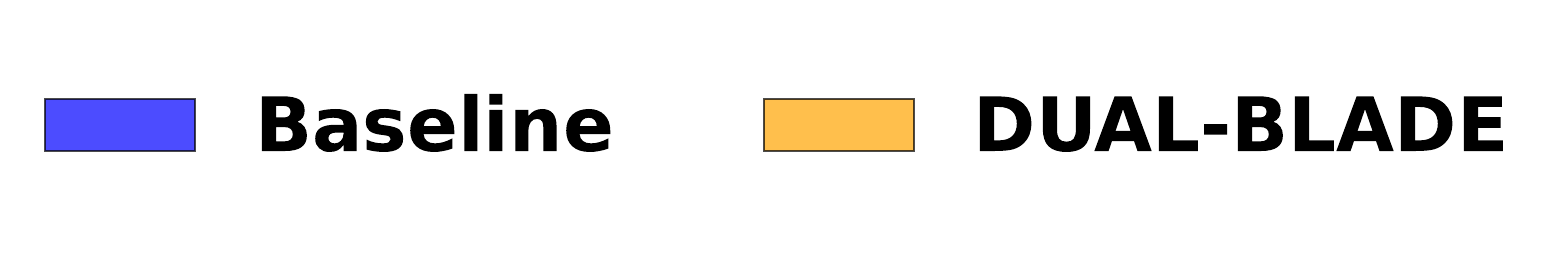}
    \vspace{-7pt} 
    \\[0pt] 
    \subfloat[\small Write (SSD~A)]{%
    \vspace{-3pt}
    \includegraphics[width=0.48\linewidth]{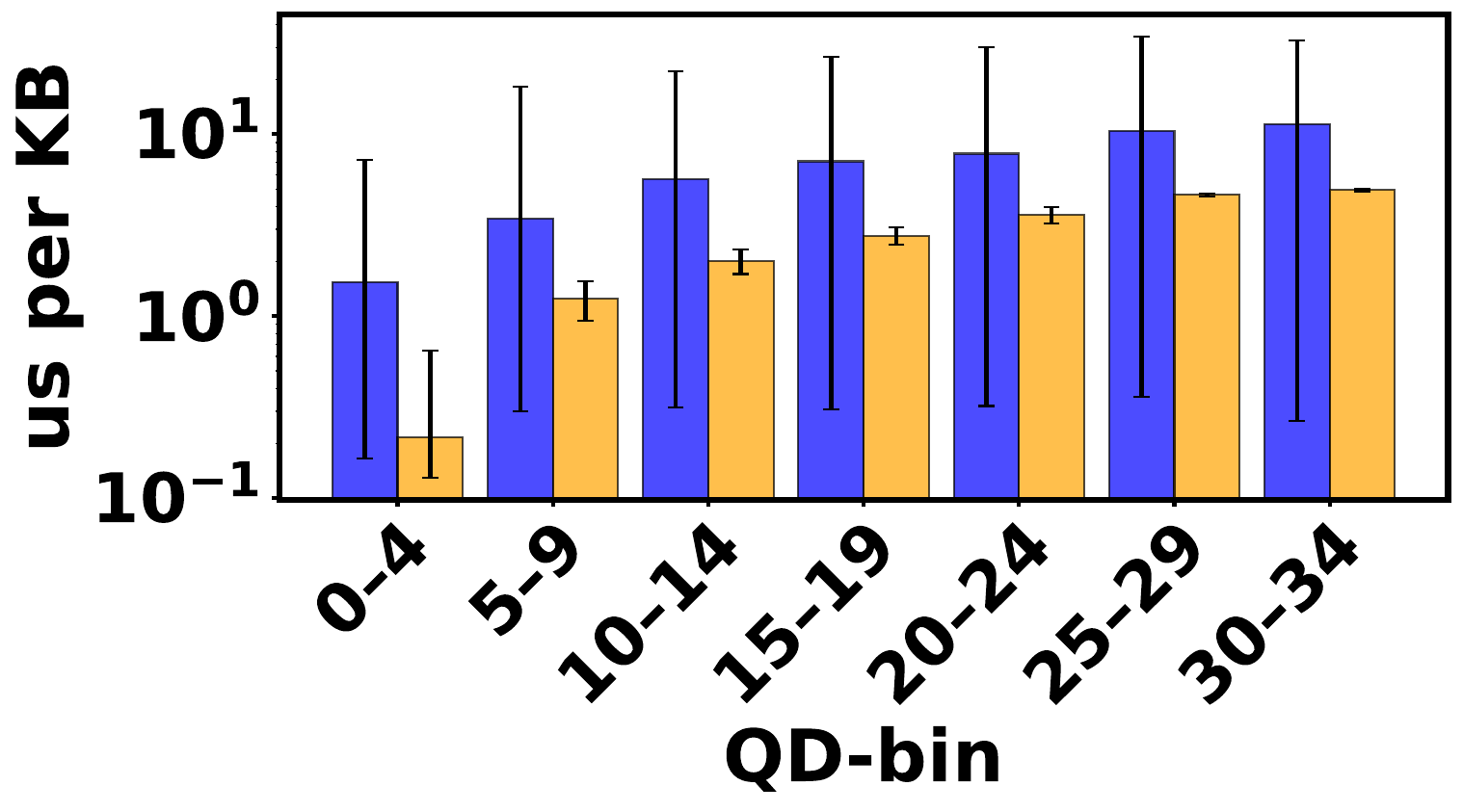}}
    \hfil
    \subfloat[\small Read (SSD~A)]{%
    \vspace{-3pt}
    \includegraphics[width=0.48\linewidth]{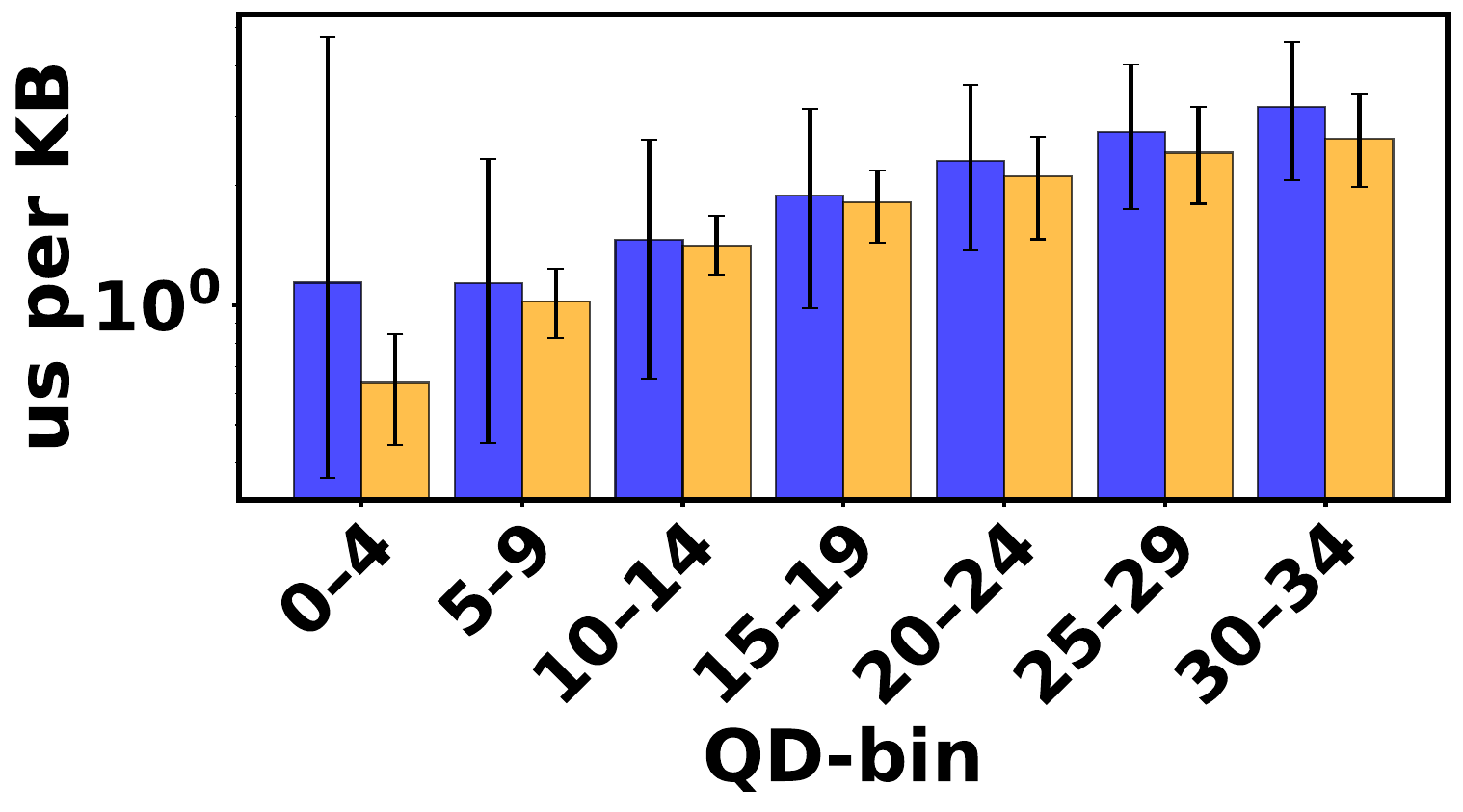}}
    \hfil
    \vspace{3pt}
    \subfloat[\small Write (SSD~B)]{%
    \vspace{-3pt}
    \includegraphics[width=0.48\linewidth]{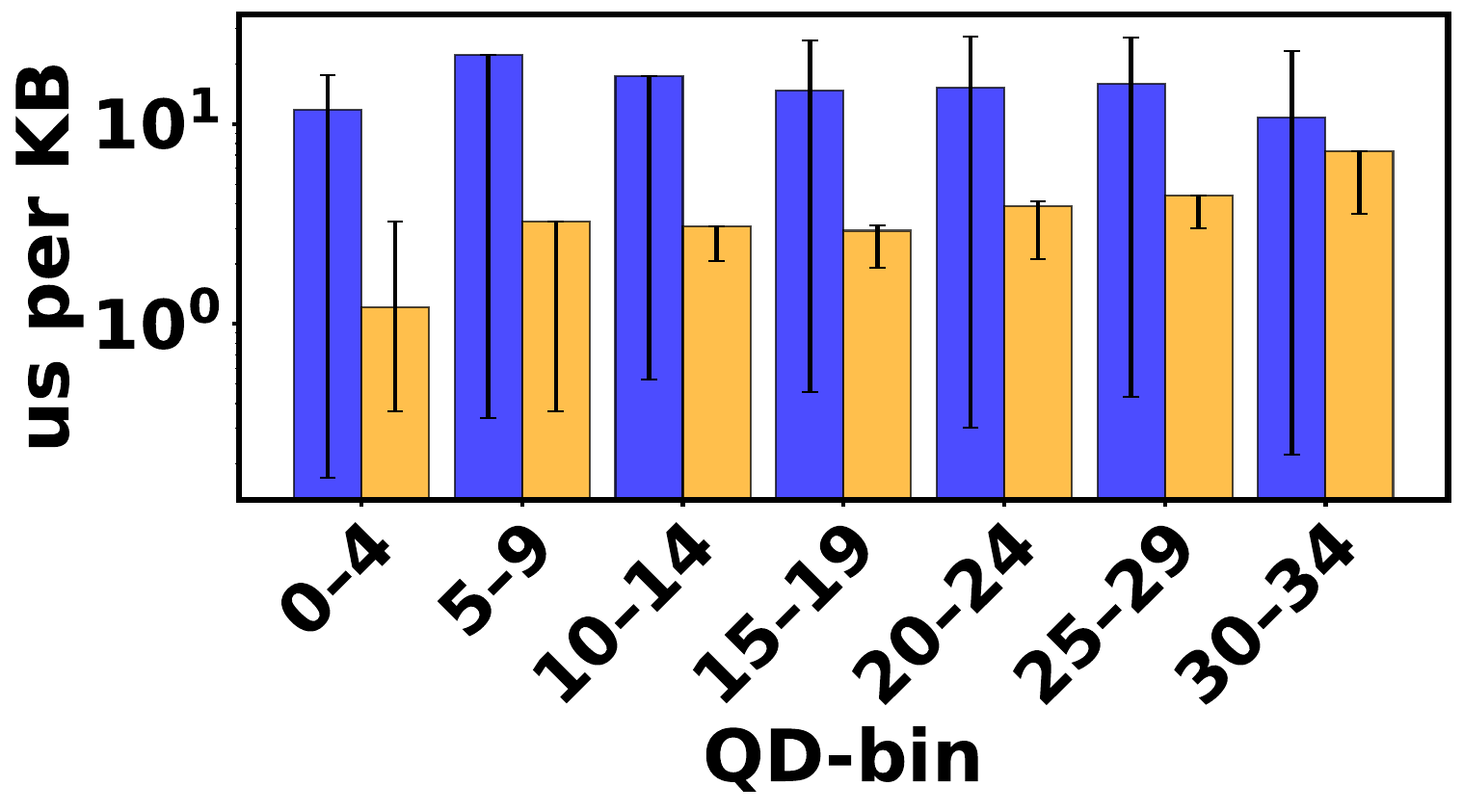}}
    \hfil
    \subfloat[\small Read (SSD~B)]{%
    \vspace{-3pt}
    \includegraphics[width=0.48\linewidth]{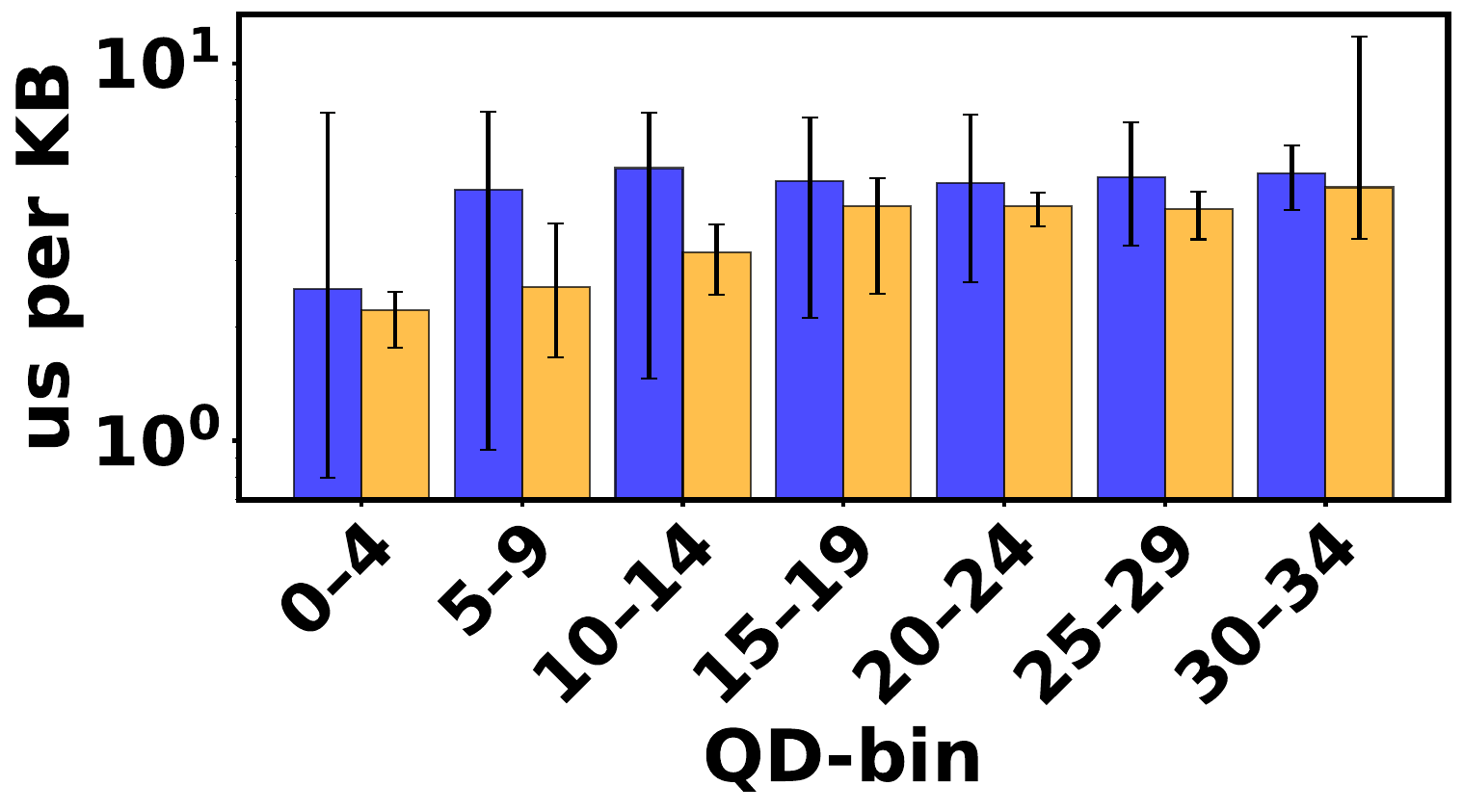}}
    \hfil
    \vspace{-3pt}
    \caption{\small Per-QD-bin I/O latency (submit$\rightarrow$complete), normalized by request size (µs/KB,  log scale). Bars show the mean. Black whiskers denote p5–p95.}
	\vspace{-3pt}
\label{fig:per_qd_bin_lat}
\end{figure}

The NVMe-direct (\S\ref{section:design_2}) further aims to sequentialize tensor access at the LBA level and reduce microsecond-scale latency inside the NVMe controller. 
To analyze this effect, we instrument NVMe submit/complete with \texttt{bpftrace} and group requests into QD bins by the observed queue depth. For a fair comparison, we cap \dualblade{} at a maximum QD of 32, and we normalize latency to μs/KB to control for request-size differences. 
Figure~\ref{fig:per_qd_bin_lat} reports per-QD-bin latency for both write and read NVMe commands on the SSD A/B.

On both SSDs, \dualblade{} lowers mean \(\mu\)s/KB and tightens the p5–p95 spread across most write/read QD bins. 
For writes, even though the Baseline shows sequential LBAs, it injects small non-sequential I/Os (e.g., journaling, metadata), which the NVMe-direct path removes. 
Similarly for reads, this design eliminates the interleaving of blk-mq streams in the Baseline.
Thus, NVMe controller processes a clean sequential stream with minimal LBA discontinuities between tensors, leading to reduced internal latency, especially at the tail.

\begin{wrapfigure}{l}{0.44\linewidth}
    \vspace{-0pt}
    \includegraphics[width=\linewidth]{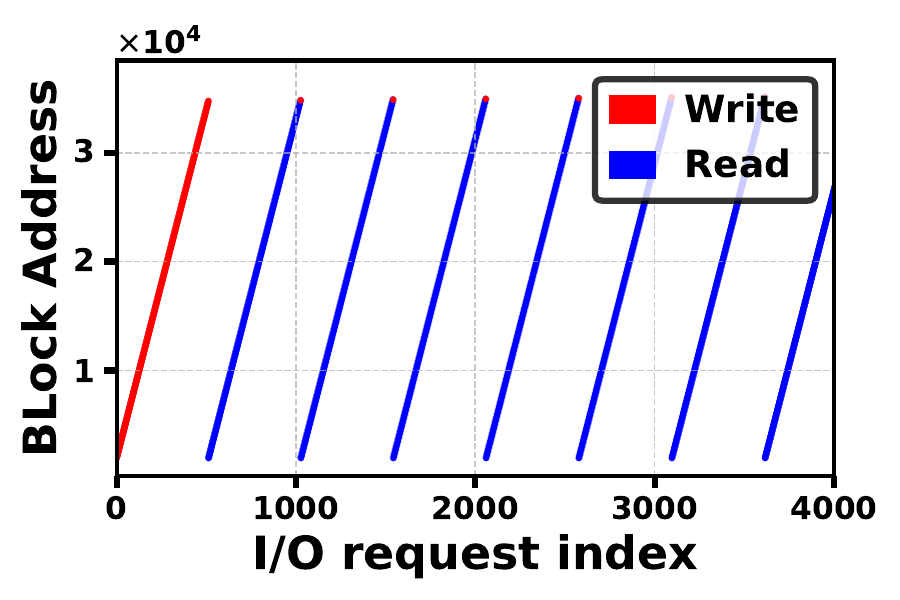}
    \caption{\small{LBA pattern.}}
    \vspace{-10pt}
    \label{fig:lba_access_pattern}
\end{wrapfigure}

Figure~\ref{fig:lba_access_pattern} plots SSD~A’s LBA access under \dualblade{} in prefill and decode, ordered by submission time, per request.
It closely matches the ideal tensor-access pattern in Figure~\ref{fig:motiv_3}(a) and, unlike the interleaved read stream of Baseline in Figure~\ref{fig:motiv_3}(b), is fully sequential, end to end. Thus the design carries the application’s sequential pattern down to device LBAs without distortion.

In a multi copy-thread environment, depending on our pipeline parallelism strategy, the I/O pattern can manifest as a single pure sequential stream or inevitably introduce the interleaving of two pure sequential streams.
However, these two pure sequential streams represent the optimal access pattern under concurrency, which modern NVMe controllers can efficiently manage without performance degradation.

\subsection{Adaptive Pipeline Parallelism}
\vspace{-2pt}
\label{sec:copy_thread_scaling}
We evaluate the impact of adaptive pipeline parallelism described in \S\ref{section:design_3} for the decode phase.
We refer to this optimization as P$\cdot$P.
We define $\alpha$ as the DRAM--SSD tiering ratio (available page-cache capacity relative to total KV size).
A higher $\alpha$ indicates that \dualblade{} serves a larger fraction of KV tensors via the page-cache path.

Table~\ref{tab:dualblade-pp-decode} compares performance with and without P$\cdot$P under $\alpha \in \{0.3, 0.5, 0.7\}$ and shows that enabling P$\cdot$P further consistently reduces decode latency across all configurations.
Across all $\alpha$, SSD~A achieves a 7--9\% latency reduction, while SSD~B shows an additional 3--5\% improvement.
This is because SSD~A's lower latency increases the relative portion of GPU DMA time, making the latency hiding achieved by overlap more effective than on the slower SSD~B.

\begin{table}[!b]
  \vspace{5pt}
  \centering
  \begin{threeparttable}
    \scriptsize
    \setlength{\tabcolsep}{2pt}
    \renewcommand{\arraystretch}{1.05}
    \caption{\small \dualblade{} decode latency with and without pipeline parallelism (P\textperiodcentered P) under varying DRAM--SSD tiering ratios $\alpha$.}
    \label{tab:dualblade-pp-decode}
    \vspace{-4pt}
    \begin{tabular*}{\linewidth}{@{\extracolsep{\fill}} l l l l l @{}}
      \toprule
      & & \multicolumn{3}{c}{Decode latency (s)} \\
      \cmidrule(lr){3-5}
      Device & Config & $\alpha=0.3$ & $\alpha=0.5$ & $\alpha=0.7$ \\
      \midrule

      \multirow{2}{*}{\textit{SSD A}} &
      \dualblade{} w/o P\textperiodcentered P & 31.13 & 28.74 & 26.59 \\
      &
      \dualblade{} w/  P\textperiodcentered P &
        \textbf{28.29~\textcolor{blue}{(\,$\times$0.91)}} &
        \textbf{26.80~\textcolor{blue}{(\,$\times$0.93)}} &
        \textbf{24.69~\textcolor{blue}{(\,$\times$0.93)}} \\

      \addlinespace[3pt]\cmidrule(lr){1-5}

      \multirow{2}{*}{\textit{SSD B}} &
      \dualblade{} w/o P\textperiodcentered P & 41.52 & 36.39 & 30.27 \\
      &
      \dualblade{} w/  P\textperiodcentered P &
        \textbf{40.09~\textcolor{blue}{(\,$\times$0.97)}} &
        \textbf{34.91~\textcolor{blue}{(\,$\times$0.96)}} &
        \textbf{28.82~\textcolor{blue}{(\,$\times$0.95)}} \\
      \bottomrule
    \end{tabular*}
  \end{threeparttable}
\end{table}

\begin{figure}[!t]
    \vspace{-10pt}
	\centering
    \includegraphics[width=0.96 \linewidth]{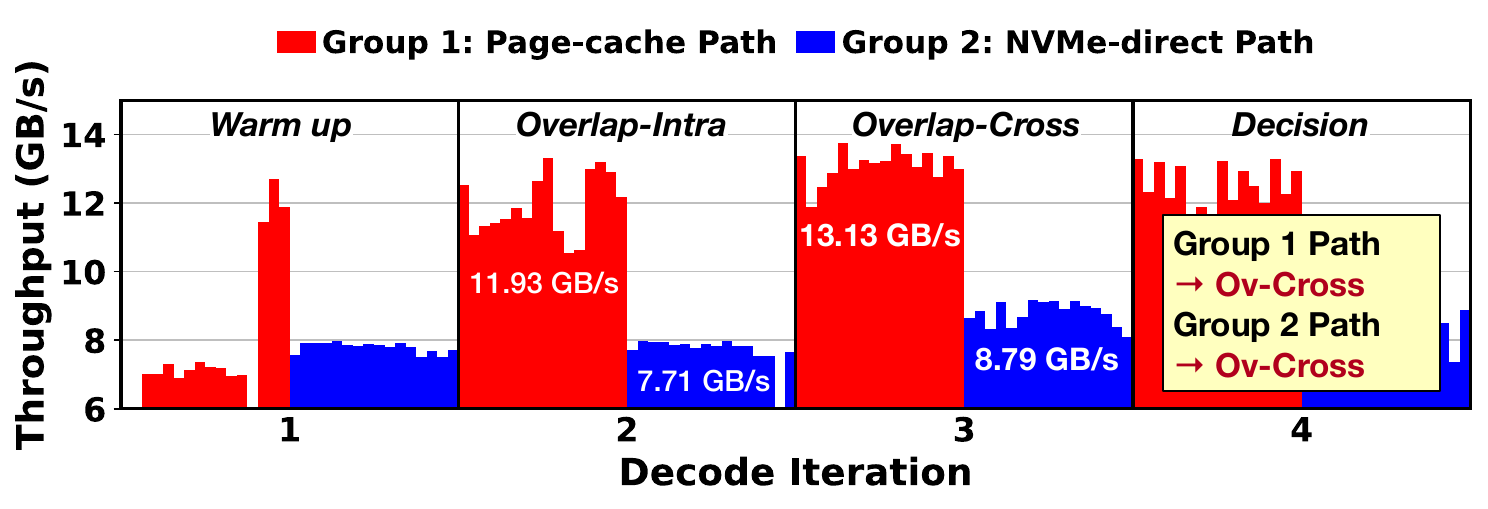}
    \vspace{-3pt}
    \caption{\small Throughput dynamics during the adaptive pipeline strategy selection on SSD A ($\alpha=0.5$).}
    \label{fig:adaptive_pp_SSD_A}
    \vspace{-5pt}
\end{figure}

To analyze \dualblade{}’s runtime strategy selection, we examine the throughput dynamics for SSD~A with $\alpha=0.5$, as shown in Figure~\ref{fig:adaptive_pp_SSD_A}.
This metric captures the per-layer throughput of KV-tensor transfers from their respective residency tiers to the GPU across decode iterations.
After a warm-up (Iteration~1), the system profiles \textit{Overlap-Intra} (Iteration~2) and \textit{Overlap-Cross} (Iteration~3).
The profiling reveals that \textit{Overlap-Cross} yields higher throughput for both groups (e.g., boosting Group~1 from 11.93 GB/s to 13.13 GB/s), leading \dualblade{} to converge on this strategy in Iteration~4.

This outcome indicates that the concurrent access in \textit{Overlap-Intra} saturates both DRAM (Group~1) and SSD (Group~2) bandwidth, resulting in device resource contention.
Conversely, \textit{Overlap-Cross} effectively mitigates this by hiding the secondary thread's I/O latency behind the primary thread's GPU DMA, maximizing end-to-end throughput.

\begin{wrapfigure}{l}{0.44\linewidth}
    \vspace{-10pt}
    \includegraphics[width=\linewidth]{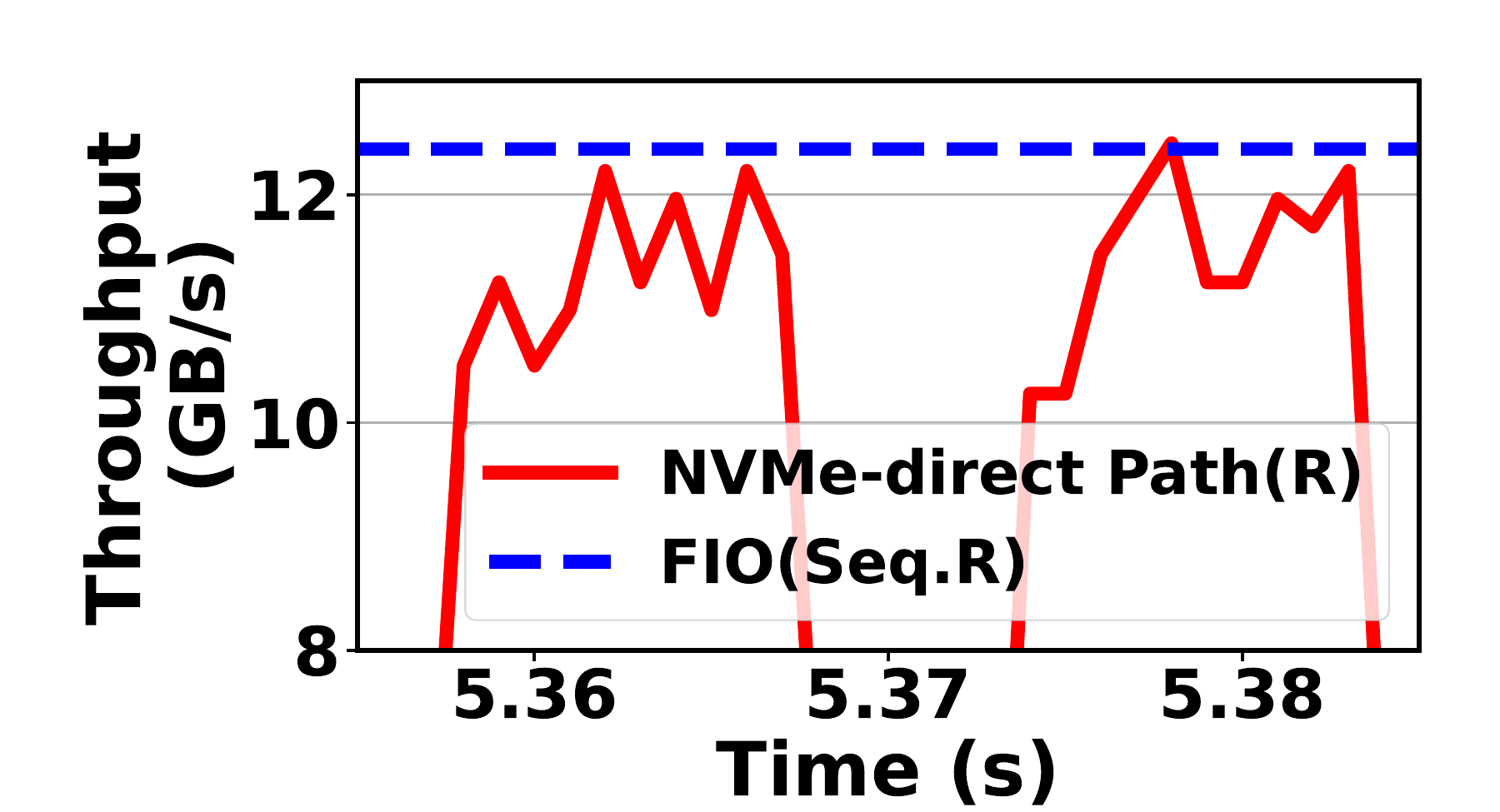}
    \caption{\small{Millisecond-level throughput analysis on SSD A. The dashed line marks the sequential read limit (FIO).}}
    \vspace{-10pt}
    \label{fig:millisencond_bw_SSD_A}
\end{wrapfigure}

To investigate the root cause of this contention, we analyzed the instantaneous throughput of a \textit{single} copy-thread (Figure~\ref{fig:millisencond_bw_SSD_A}).
Unlike the per-second average in Figure~\ref{fig:ssd_throughput}, this fine-grained analysis at millisecond resolution captures the instantaneous throughput spikes, revealing that a single copy-thread alone is sufficient to saturate the SSD's peak sequential read bandwidth.
Thus, the concurrent reads from multi copy-threads in \textit{Overlap-Intra} coincided within these short, saturated intervals, resulting in device resource contention.

\subsection{Evaluation on Practical Edge Scenarios}
\vspace{-2pt}
\label{sec:copy_thread_scaling}
Beyond interactive chat applications, Edge AI systems utilize LLMs for \textit{back-of-house} data wrangling to continuously clean and structure raw operational data~\cite{zhong2024logparser, wang2025intelligent}. 
Specifically, they execute Entity Matching (EM), Data Imputation (DI), and Error Detection (ED) at the source.

As shown in Table~\ref{tab:data_wrangle_result}, we evaluate four representative data wrangling tasks~\cite{narayan2022can} with OPT-6.7B.
These tasks are characterized by long input contexts (200--744 tokens) and short outputs (3--10 tokens).
We conducted the experiments with a batch size of 32 under a strict 4~GB host memory limit. This leaves much less than 4 GB for the KV-cache.

\dualblade{} outperforms the Baseline in most cases (e.g., reducing latency to $0.85\times$ in DI:Buy). 
An exception is ED:Hospital, where performance remains comparable ($1.00\times$) because the relatively small KV-cache (1.58 GB) fits entirely within the available page-cache.

\section{Related Work}
\vspace{-2pt}
\label{sec:related}

\noindent\textbf{SSD-Enabled LLM Inference Techniques.}
Studies have explored offloading techniques for LLM inference to NVMe SSDs to overcome memory limits.
LLM-in-a-Flash~\cite{alizadeh2024llm} achieves weight-centric offloading by storing model parameters on flash for on-demand loading via flash-friendly transfers.
KVSwap~\cite{zhang2025kvswap} uses a low-rank predictor to approximate attention scores, selectively fetching critical KV groups from flash storage.
AttentionStore~\cite{gao2024cost} enables multi-turn reuse by persisting KV states across a DRAM--SSD hierarchy for efficient reinstatement.
A quantitative study~\cite{ren2025characterizing} identifies NVMe-level I/O bottlenecks when offloading model weights or KV tensors across inference frameworks~\cite{sheng2023flexgen, aminabadi2022deepspeed}. InstInfer~\cite{pan2024instinfer} and INF$^2$~\cite{jang2025inf} target Computational Storage Devices (CSDs) over commodity SSDs to offload attention/KV handling in-storage.

At the system layer, LLM serving frameworks like FlexLLMGen~\cite{sheng2023flexgen}, llama.cpp~\cite{llama_cpp}, and vLLM~\cite{kwon2023efficient} (with LMCache~\cite{cheng2025lmcache}) leverage heterogeneous memory tiers spanning GPU memory, host DRAM, and SSDs to manage model weights and KV tensors.
\dualblade{} is orthogonal to FlexLLMGen and can work in conjunction with it to offload the KV-cache to SSD more efficiently. 
Its design principles are framework-agnostic and can be applied to KV offloading backends (e.g., LMCache) to accelerate storage I/O across different serving stacks.

\vspace{1pt}
\noindent\textbf{Hardware-Oriented Data-Movement Mechanisms.} 
Beyond LLM-specific offloading policies, several hardware-oriented primitives define the data path among GPU memory, host DRAM, and storage. 
UVM~\cite{nvidia_cuda_pg_unified_system_mem} provides a unified managed address space and on-demand page migration across CPU and GPU, enabling oversubscription beyond physical VRAM by leveraging host DRAM, which is useful on low-memory GPUs.
GDS~\cite{nvidia_gds_online} enables direct storage-to-GPU DMA by bypassing the CPU staging path (host bounce buffers), supporting GPU-targeted I/O directly into device memory.
BaM~\cite{qureshi2023gpu} enables GPU-initiated direct storage-to-GPU access via GPU-resident submission/completion queues (doorbells), supporting fine-grained transfers with minimal CPU involvement.

Although GDS and BaM can accelerate storage offloading via GPU--SSD P2P DMA, these capabilities typically assume datacenter-class GPUs (e.g., A100/H100) and supported server platforms with compatible PCIe topologies, making them best suited for cloud-scale inference deployments. 
In contrast, \dualblade{} targets edge AI platforms (e.g., on-premise and embedded deployments) where GPU--SSD P2P is unavailable.
Furthermore, \dualblade{}’s designs can be integrated on top of GDS/BaM-style GPU--SSD P2P primitives when available, further improving offloading performance.

\begin{table}[!t]
  \centering
  \begin{threeparttable}
    \scriptsize
    \setlength{\tabcolsep}{2pt}
    \renewcommand{\arraystretch}{1.05}
    \caption{\small Inference latency (s) of data wrangling tasks~\cite{narayan2022can}.}
    \label{tab:data_wrangle_result}
    \vspace{-4pt}
    \begin{tabular*}{\linewidth}{@{\extracolsep{\fill}} l c c c c c c @{}}
      \toprule
      & & & \multicolumn{2}{c}{\textit{SSD A}} & \multicolumn{2}{c}{\textit{SSD B}} \\
      \cmidrule(lr){4-5}\cmidrule(lr){6-7}
      Dataset & Queries & \begin{tabular}[c]{@{}c@{}}KV-cache\\(GB)\end{tabular}
              & Base & \dualblade{}
              & Base & \dualblade{} \\
      \midrule
      EM:Fodors-Zagats  & 189 & 5.84 & 75.28 &
        \textbf{67.17}~\textbf{\textcolor{blue}{($\times$0.89)}} &
        86.40 & \textbf{79.06}~\textbf{\textcolor{blue}{($\times$0.92)}} \\
      EM:Walmart-Amazon & 200 & 5.87 & 80.92 &
        \textbf{72.18}~\textbf{\textcolor{blue}{($\times$0.89)}} &
        93.17 & \textbf{85.41}~\textbf{\textcolor{blue}{($\times$0.92)}} \\
      DI:Buy            &  65 & 3.89 & 30.55 &
        \textbf{27.90}~\textbf{\textcolor{blue}{($\times$0.91)}} &
        34.05 & \textbf{29.27}~\textbf{\textcolor{blue}{($\times$0.86)}} \\
      ED:Hospital       & 200 & 1.58 & 19.63 &
        \textbf{19.60}~\textbf{\textcolor{blue}{($\times$1.00)}} &
        19.62 & \textbf{19.66}~\textbf{\textcolor{blue}{($\times$1.00)}} \\
      \bottomrule
    \end{tabular*}
  \end{threeparttable}
\end{table}

\section{Conclusion}
\label{sec:conclusion}
\vspace{-2pt}
This paper presents \dualblade{}, a dual-path, NVMe-direct KV-cache offloading architecture to accelerate LLM inference in edge AI systems with limited memory.
\dualblade{} integrates dual-path KV residency, NVMe-direct I/O, and adaptive pipeline parallelism to maximize storage I/O efficiency during inference.
Extensive evaluation shows that \dualblade{} reduces prefill and decode latency by up to 33.1\% and 42.4\%, respectively, while boosting SSD utilization by up to 2.2$\times$ in diverse memory budgets.

\def\IEEEbibitemsep{0.15pt plus 0pt} 
\renewcommand{\baselinestretch}{0.9} 

\bibliographystyle{ieeetr}
\bibliography{ref} 

\end{document}